\newif\ifconfver
\confvertrue        

\newif\ifplainver  

\ifplainver
    \confverfalse   
\fi

\ifconfver
     \documentclass[10pt,twocolumn,twoside]{IEEEtran}
\else
    \ifplainver
        \documentclass[11pt]{article}
        \usepackage{fullpage}
    \else
        \documentclass[11pt,draftcls,onecolumn]{IEEEtran}
    \fi
\fi
\usepackage{calc,amsfonts,amssymb,amsmath,bm,url,color,theorem,graphicx,cite}
\usepackage{psfrag,subfigure,float}
\usepackage{algorithm}
\usepackage{algorithmic}

\definecolor{orange}{RGB}{255,107,0}

\newtheorem{Lemma}{Lemma}
\newtheorem{Prop}{Proposition}
\newtheorem{Theorem}{Theorem}
\newtheorem{Assumption}{Assumption}
\newtheorem{Def}{Definition}
\newtheorem{Corollary}{Corollary}

\newtheorem{Observation}{Observation}
\theorembodyfont{\rmfamily}

\newtheorem{Remark}{Remark}

\begin{document}

\bibliographystyle{IEEEtran}

\newcommand{\papertitle}{
Active Sensing of Social Networks
}

\newcommand{\paperabstract}{
...
}


\ifplainver

    \title{\papertitle}

    \author{
    Hoi-To Wai, Anna Scaglione, Amir Leshem
    }

    \maketitle

\else
    \title{\papertitle}

    \ifconfver \else {\linespread{1.1} \rm \fi

    \author{Hoi-To Wai,~\emph{Student Member, IEEE,} Anna Scaglione,~\emph{Fellow, IEEE,} and Amir Leshem,~\emph{Senior Member, IEEE}
    \thanks{This material is based upon work supported by NSF CCF-1011811 and partially supported by ISF 903/13. A preliminary version of this work has been presented at IEEE CDC 2015, Osaka, Japan \cite{cdc_submit}.}
        \thanks{H.-T. Wai and A. Scaglione are with the School of Electrical, Computer and Energy Engineering, Arizona State University, Tempe, AZ 85281, USA. E-mails: \texttt{\{htwai,Anna.Scaglione\}@asu.edu}. A. Leshem is with Faculty of Engineering, Bar-Ilan University, Ramat-Gan, Israel. Email: \texttt{leshema@eng.biu.ac.il}}}

    \maketitle

    \ifconfver \else
        \begin{center} \vspace*{-2\baselineskip}
        \end{center}
    \fi

    \begin{abstract}
This paper develops an active sensing method to estimate the relative weight (or trust) agents place on their neighbors' information in a social network. The model used for the regression is based on the steady state equation in the linear DeGroot model under the influence of stubborn agents; i.e., agents whose opinions are not influenced by their neighbors. This method can be viewed as a \emph{social RADAR}, where the stubborn agents excite the system and the latter can be estimated through the reverberation observed from the analysis of the agents' opinions. 
{The social network sensing problem can be interpreted as a blind compressed sensing problem with a sparse measurement matrix. We prove that the network structure will be revealed when a sufficient number of stubborn agents independently influence a number of ordinary (non-stubborn) agents.}
We investigate the scenario with a deterministic or randomized DeGroot model and propose a consistent estimator of the steady states for the latter scenario. 
{Simulation results on synthetic and real world networks support our findings.}
    \end{abstract}

    \begin{keywords}\vspace{-0.0cm}
       DeGroot model, opinion dynamics, social networks, sparse recovery, system identification
    \end{keywords}

    \ifconfver \else \IEEEpeerreviewmaketitle} \fi

 \fi

\ifconfver \else
    \ifplainver \else
        \newpage
\fi \fi
\section{Introduction}
\IEEEPARstart{R}ecently, the study of networks has become a major  research focus in many disciplines, where \emph{networks} have been used to model systems from biology, physics to the social sciences \cite{Jackson2008}.
From a signal processing perspective, the related research problems can be roughly categorized into \emph{analysis, control and sensing}. While these are natural extensions of classical signal processing problems, the spatial properties due to the underlying network structure have yielded new insights and fostered the development of novel signal processing tools \cite{Sayed2013,Anis2014,Sandryhaila2014}.

The \emph{network analysis} problem has received much attention   due to the emergence of `big-data' from social networks, biological networks, etc. Since the networks to be analyzed consist of millions of vertices and edges, computationally efficient tools have been developed to extract low-dimensional structures, e.g., by detecting community structures \cite{Clauset2004,Fortunato2010} and finding the centrality measures of the vertices \cite{page1999pagerank}. Techniques in the related studies involve developing tools that run in (sub)-linear time with the size of the network; e.g., see \cite{Ishii2014}.
Another problem of interest is  known as \emph{network control}, where the  aim is to choose a subset of vertices that can provide full/partial control over the  network. It was shown recently in \cite{Liu2011} that the Kalman's classical control criterion is equivalent to finding a maximal matching on the network; see also \cite{Liu2013,Doostmohammadian2014}. 

This work considers the \emph{social network sensing} problem that has received relatively less attention in the community than the two aforementioned problems. We focus on social networks\footnote{  That said, the developed methodology is general and may be applied to other types of networks with similar dynamics models.} and our aim is to \emph{infer} simultaneously the trust matrix and the network structure  by observing the opinion dynamics. We model the \emph{opinion dynamics} in the social network according to the DeGroot model \cite{degroot}. In particular, at each time step, the opinions are updated by taking a convex combination of the neighboring opinions, weighted with respect to a \emph{trust matrix}. 
{  
Despite its simplicity, the DeGroot model has been widely popular in the social sciences; some experimental papers indicated that the model is able 
 to capture  the actual opinion dynamics, e.g.,  \cite{De2014,das2014modeling,Jia2013,chandra2012,acemoglu2011opinion}.}
Notice that the DeGroot model is analogous to the Average Consensus Gossiping (ACG) model \cite{Blondel2005,Xiao2007} and drives the opinions of every agent in the network to \emph{consensus} as time goes to infinity.
Hence, in such situations the social network sensing method is required to \emph{track} the individual opinion updates. This is the \emph{passive} approach taken in previous works \cite{Timme2007,Wang2011b,De2014,Mei2015}. 
As agents' interactions may occur asynchronously and at an unknown timescale, these approaches may be impractical due to the difficulty in precisely tracking the opinion evolution.

An interesting extension in the study of opinion dynamics is to consider the effects of stubborn agents (or zealots), whose opinions remain unchanged throughout the network dynamics. 
Theoretical studies have focused on {  characterizing the steady-state of the opinion dynamics when stubborn agents are present} \cite{yildiz2011discrete,yildiz2013binary,Acemoglu2013,Jia2013,mauro03,alex15,yildiz2010computing}, developing techniques to effectively \emph{control} the network and attain fastest convergence \cite{yildiz2010computing,bianchi2012,Khan2010,Wang2014}. 
{The model of stubborn agents has also been studied in the context of sensor network 
localization \cite{khan09}.}
{More recently, experimental studies have provided evidence supporting the existence of stubborn agents
in social networks, for instance, \cite{das2014modeling,mehdi13} suggested that stubborn agents 
can be used to justify several opinion dynamic models for data collected from controlled experiments; 
\cite{marlon15} illustrated that the existence of stubborn agents is a plausible cause for the emergence 
of extreme opinions in social networks; \cite{kuhlman12} studied the controllability of opinions in 
real social networks using stubborn agents.}

{The aim of this paper is to demonstrate an important consequence of the existence of \emph{stubborn agents}. 
Specifically, we propose a \emph{social RADAR} formulation through exploiting the special role of stubborn agents.
As will be shown below, the stubborn agents will help expose the network structure through a set of steady state equations.}
The proposed model shows that the stubborn agents can play a role that is analogous to a traditional radar, which \emph{actively} scans and probes the social network.

As the stubborn agents may constitute a small portion of the network, the set of steady state equation may be an undetermined system. To handle this issue, we formulate the network sensing problem as a sparse recovery problem using the popular $\ell_0/\ell_1$ minimization technique. Finally, we derive a low-complexity optimization method for social network sensing. The proposed method is applicable to large networks.
An additional contribution of our investigation is a new recoverability result for a special case of \emph{blind compressive sensing}; e.g.,\cite{Gleichman2011,Studer2012}. In particular, we develop an identifiability condition for active sensing of social networks based on expander graph theories \cite{Gilbert2010}. Compared to \cite{Berinde2008,Wang2011,Khajehnejad2011,Gilbert2010}, our result is applicable when the sensing matrix is non-negative (instead of binary) and subject to small multiplicative perturbations.

The remainder of this paper is organized as follows. Section~\ref{sec:sys} introduces the DeGroot model for opinion dynamics and the effects of stubborn agents. Section~\ref{sec:netwk} formulates the social network sensing problem as a sparse recovery problem and provides a sufficient condition for perfect recovery. Section~\ref{sec:fast} further develops a fast optimization method for approximately solving the sparse recovery problem and {  shows that the developed methods can be applied when the opinion dynamics is a time-varying system}. Simulation results on both synthetic  and a real network example conclude the paper, in Section~\ref{sec:sim}.

\textbf{\emph{Notation:}} We use boldfaced small/capital letters to denote vectors/matrices and $[n] = \{1,...,n\}$ to represent the index set from $1$ to $n$ and $n \in \mathbb{N}$.
${\rm Diag}: \mathbb{R}^n \rightarrow \mathbb{R}^{n \times n}$/${\rm diag}: \mathbb{R}^{n \times n} \rightarrow \mathbb{R}^n$ to denote the diagonal operator that maps from a vector/square matrix to a square matrix/vector, $\geq$ to represent the element-wise inequality. For an index set $\Omega \subseteq [n]$, we use $\Omega^c$ to denote its natural complement, e.g., $\Omega^c = [n] \setminus \Omega$. A vector ${\bm x}$ is called $k$-sparse if $\| {\bm x} \|_0 \leq k$. We denote ${\rm off}({\bm W})$ as the square matrix with only off-diagonal elements in ${\bm W}$.

\section{Opinion Dynamics Model} \label{sec:sys}
Consider a social network represented by a simple, connected weighted directed graph $G = (V,E,\overline{\bm W})$, where $V = [n]$ is the set of agents, $E \subseteq V \times V$ is the network topology and $\overline{\bm W} \in \mathbb{R}^{n \times n}$ denotes the \emph{trust} matrix between agents. Notice that $\overline{\bm W}$ can be asymmetric. The trust matrix satisfies $\overline{\bm W} \geq {\bf 0}$ and $\overline{\bm W}{\bf 1} = {\bf 1}$, i.e., $\overline{\bm W}$ is stochastic. Furthermore, $\overline{W}_{ii} < 1$ for all $i$ and $\overline{W}_{ij} > 0$ if and only if $ij \in E$.

We focus on the linear DeGroot model \cite{degroot} for opinion dynamics. 
{  There are $K$ different discussions  in the social network and each discussion is indexed by $k \in [K]$}.
Let $x_i(t;k)$ be the opinion of the $i$th agent\footnote{  For example, the $i$th agent's opinion $x_i(t;k)$ may represent a probability distribution function of his/her stance on the discussion. While our discussion is focused on the case when $x_i(t;k)$ is a scalar, it should be noted that the techniques developed can be easily extended to the vector case; see \cite{cdc_submit}.} at time $t \in \mathbb{N}$ during the $k$th {  discussion}.  
{ 
Using the intuition that individuals' opinions are influenced by opinions of the others, the DeGroot model postulates that}  the agents' opinions are updated according to the following random process:
\begin{equation} \label{eq:dyn}
{x}_i(t;k) = \sum_{j \in {\cal N}_i} W_{ij}(t;k) x_j(t-1;k), 
\end{equation}
which can also be written as
\begin{equation} \label{eq:dyn_mat}
{\bm x}(t;k) = {\bm W}(t;k) {\bm x}(t-1;k),\end{equation}
 where ${\bm x}(t;k) = (x_1(t;k),\ldots,x_n(t;k))^T$, $[{\bm W}(t;k)]_{ij} = {W}_{ij}(t;k)$ and ${\bm W}(t;k)$ is non-negative and stochastic, i.e., ${\bm W}(t;k) {\bf 1} = {\bf 1}$ for all $t,s$.
{  See \cite{degroot} and \cite[Chapter 1]{friedkin11} for a detailed description of the aforementioned model.}
We assume the following regarding the random matrix ${\bm W}(t;k)$. 
\begin{Assumption} {  ${\bm W}(t;k)$ is an independently and identically distributed (i.i.d.)~random matrix drawn from a distribution that satisfies $\mathbb{E} \{ {\bm W}(t;k) \} = \overline{\bm W}$ for all $t \in \mathbb{N}, k \in [K]$. }
\end{Assumption}

The agents' opinions can be observed as:
\begin{equation}\label{eq:obs}
\hat{\bm x}(t;k) = {\bm x}(t;k) + {\bm n}(t;k),
\end{equation}
where ${\bm n}(t;k)$ is i.i.d.~additive noise with zero-mean and bounded variance.
Eq.~\eqref{eq:dyn} \& \eqref{eq:obs} constitute a time-varying linear dynamical system.
Since $\overline{\bm W}$ encodes both the network topology and the trust strengths, the social network sensing problem is to infer $\overline{\bm W}$ through the set of observations $\{ \hat{\bm x}(t;k) \}_{t \in {\cal T},k \in [K]}$, where ${\cal T}$ is the set of sampling instances.

A well known fact in the distributed control literature \cite{Blondel2005} is that the agents' opinions in Eq.~\eqref{eq:dyn} attain consensus almost surely as $t \rightarrow \infty$:
\begin{equation} \label{eq:asymp_norm}
\lim_{ t \rightarrow \infty } {\bm x}(t;k) =^{a.s.} {\bf 1} {\bm c}^T(s) {\bm x}(0;k),
\end{equation}
for some ${\bm c}(s) \in \mathbb{R}^n$, i.e., $x_i(t;k) = x_j(t;k)$ for all $i,j$ as $t \rightarrow \infty$. In other words, the information about the network structure vanishes in the steady state equation.

The aforementioned property of the DeGroot model has prompted most works on social network sensing (or network reconstruction in general) to assume a complete/partial knowledge of ${\cal T}$ such that the opinion dynamics can be tracked.
  In particular, \cite{De2014} assumes a static model such that ${\bm W}(t;k) = \overline{\bm W}$ and infers $\overline{\bm W}$ by solving a least square problem; \cite{Timme2007,Wang2011b} deals with a time-varying, non-linear dynamical system model and applies sparse recovery to infer $\overline{\bm W}$.  The drawback of these methods is that they require knowing the discrete time indices for the observations made.
This knowledge may be difficult to obtain in general.
In contrast, the actual system states are updated with an \emph{unknown} interaction rate and the interaction timing between agents is not observable in most practical scenarios. 

Our active sensing method relies on observing the steady state opinions; i.e., the opinions when $t \rightarrow \infty$. The observations are based on ${\cal T}$ such that $\min (t \in {\cal T}) \gg 0$ and are thus robust to errors in tracking of the discrete time dynamics.
Our approach is to reconstruct the network via a set of \emph{stubborn agents}, as described next.

\subsection{Stubborn Agents (a.k.a.~zealots)}
Stubborn agents (a.k.a.~zealots) are members whose opinions can not be swayed by others. {  If agent $i$ is stubborn, then $x_i(t;k) = x_i(0;k)$ for all $t$}. The opinion dynamics of these agents can be characterized by the structure of their respective rows in the trust matrix:
\begin{Def}
An agent $i$ is stubborn if and only if its corresponding row in the trust matrix ${\bm W}(t;k)$  is the canonical basis vector; i.e., for all $j$,
\begin{equation}
{W}_{ij} (t;k) = \begin{cases}
1,~{\rm if}~{j = i}, \\
0,~{\rm otherwise},
\end{cases}
\forall~t,k
\end{equation}
\end{Def}
{  Note that stubborn agents are known to exist in social networks, as discussed in the Introduction.}

Suppose that there exists $n_s$ stubborn agents in the social network and the \emph{social RADAR} is aware of their existence. 
Without loss of generality, we let $V_s \triangleq [n_s]$ be the set of stubborn agents and $V_r \triangleq V \setminus V_s$ be the set of ordinary agents. 
The trust matrix $\overline{\bm W}$ and its realization can be written {  as the following block matrices:}
\begin{equation}
\overline{\bm W} = \left(
\begin{array}{cc}
{\bm I}_{n_s} & {\bf 0} \\
\overline{\bm B} & \overline{\bm D}
\end{array}
\right),~
{\bm W} (t;k) = \left(
\begin{array}{cc}
{\bm I}_{n_s} & {\bf 0} \\
{\bm B}(t;k) & {\bm D}(t;k)
\end{array}
\right),
\end{equation}
{  where ${\bm B}(t;k)$ and ${\bm D}(t;k)$ are the partial matrices of ${\bm W}(t;k)$,}
note that $\mathbb{E} \{ {\bm B} (t;k) \} = \overline{\bm B}$ and $\mathbb{E} \{ {\bm D} (t;k) \} = \overline{\bm D}$. {  Notice that $\overline{\bm B}$ is the network between stubborn and ordinary agents, and $\overline{\bm D}$ is the network among the ordinary agents themselves.}

We further impose the following assumptions:

\begin{Assumption} \label{assume_1}
The support of $\overline{\bm B}$, $\Omega_{\overline{\bm B}} = \{ ij : \overline{B}_{ij} > 0 \} = E(V_r,V_s)$, is known. Moreover, each agent in $V_r$ has non-zero trust in at least one agent in $V_s$.
\end{Assumption}
\begin{Assumption} \label{assume_2}
The induced subgraph $G[V_r] = (V_r, E(V_r))$ is connected. 
\end{Assumption}
Consequently, the principal submatrix $\overline{\bm D}$ of $\overline{\bm W}$ satisfies $\| \overline{\bm D} \|_2 < 1$.

We are interested in the steady state opinion resulting from \eqref{eq:dyn} at $t \rightarrow \infty$.
\begin{Observation} \label{obs:s} \cite{khan09,yildiz2010computing} Let ${\bm x}(t;k) \triangleq ( {\bm z}(t;k),~{\bm y}(t;k) )^T \in \mathbb{R}^{n}$ where ${\bm z}(t;k) \in \mathbb{R}^{n_s}$ and ${\bm y}(t;k) \in \mathbb{R}^{n-n_s}$ are the opinions of stubborn agents and ordinary agents, respectively. Consider \eqref{eq:dyn} by setting $t \rightarrow \infty$, we have:
\begin{equation}
\lim_{t \rightarrow \infty} \mathbb{E} \{ {\bm x}(t;k) | {\bm x} (0;k) \} = \overline{\bm W}^\infty {\bm x} (0;k),
\end{equation}
where
\begin{equation} \label{eq:stubborn_w}
\overline{\bm W}^\infty = \left( \begin{array}{cc}
{\bm I}_{n_s} & {\bm 0} \\
( {\bm I} - \overline{\bm D})^{-1} \overline{\bm B} & {\bm 0}
\end{array}\right). \vspace{.2cm}
\end{equation}
\end{Observation}
The expected opinions of ordinary agents at $t \rightarrow \infty$ are:
\begin{equation} \label{eq:stubborn_asy}
\lim_{t \rightarrow \infty} \mathbb{E} \{ {\bm y} (t;k) | {\bm z}(0;k) \} = ( {\bm I} - \overline{\bm D})^{-1} \overline{\bm B} {\bm z}(0;k).
\end{equation}
As seen, the steady state opinions are dependent on the stubborn agents and the structure of the network.
Furthermore, unlike the case \emph{without} stubborn agents (cf.~\eqref{eq:asymp_norm}), information about the network structure $\overline{\bm D}, \overline{\bm B}$ will be retained in the steady state equation \eqref{eq:stubborn_asy}. Leveraging on Observation~\ref{obs:s}, the next section formulates a regression problem that estimates $\overline{\bm D}, \overline{\bm B}$ from observing opinions that fit these steady state equations.

\section{Social Network Sensing via Stubborn Agents} \label{sec:netwk}
\begin{figure}[t]
\centering
 \includegraphics[width =0.75\linewidth]{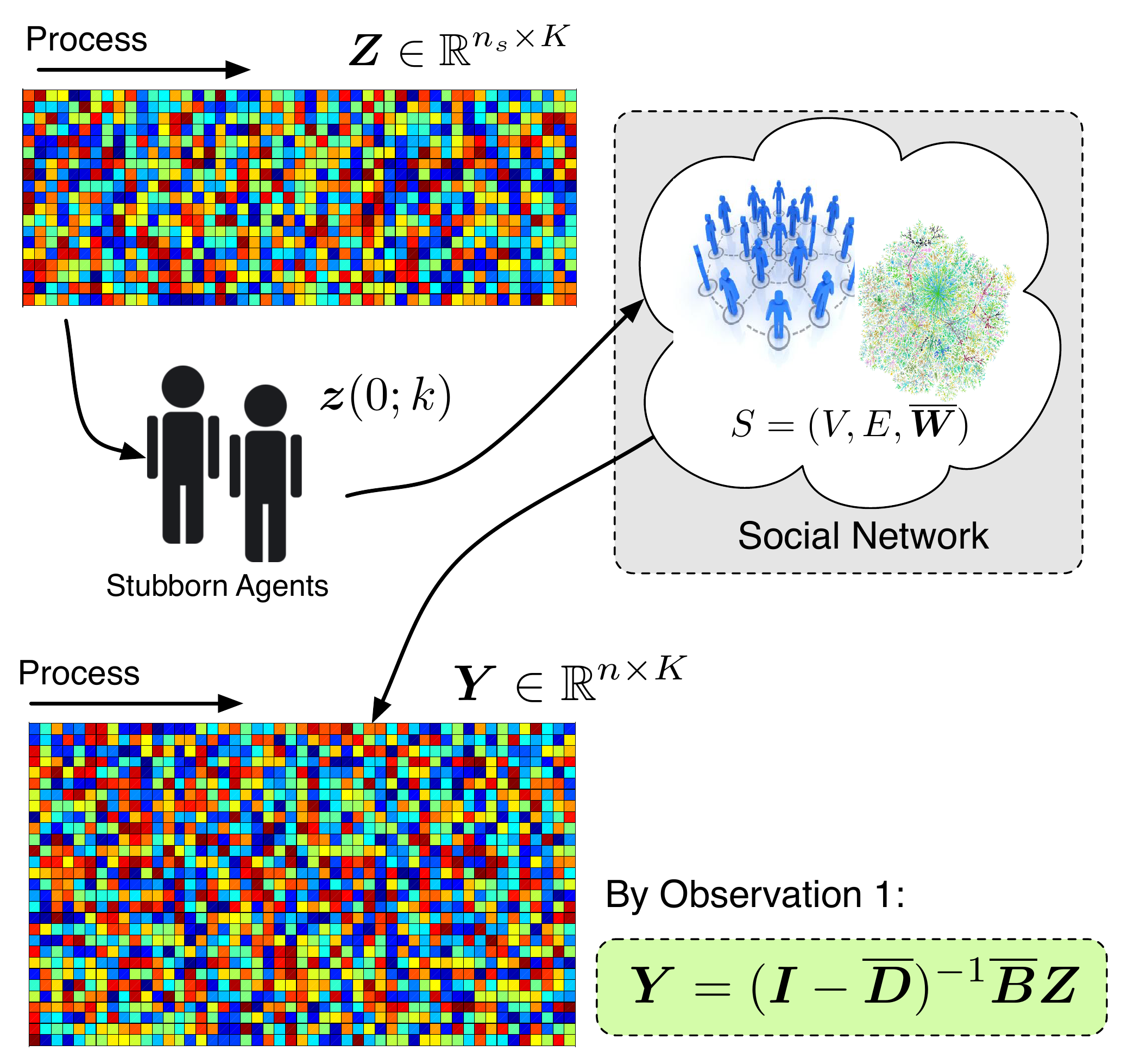}
\caption{Data matrices used in the social network sensing problem (cf. \eqref{eq:collect}).} \vspace{-.2cm}
\label{fig:col}
\end{figure}

We now study the problem of social network sensing using stubborn agents.
{  Instead of tracking the opinion evolution in the social networks similar to the \emph{passive} methods \cite{De2014,Timme2007,Wang2011b}, our method is based on  collecting the steady-state opinions from $K \geq n_s$ discussions.}
Define the data matrices:
\begin{subequations} \label{eq:collect}
\begin{align}
{\bm Y} \triangleq & \Big( \mathbb{E}\{{\bm y}(\infty;1)\}~\cdots~ \mathbb{E}\{{\bm y}(\infty;K)\} \Big) \in \mathbb{R}^{(n-n_s) \times K}, \\
{\bm Z} \triangleq & \Big( {\bm z}(0;1)~\cdots~ {\bm z}(0;K) \Big)  \in \mathbb{R}^{n_s \times K}.
\end{align}
\end{subequations}
Notice that \eqref{eq:stubborn_asy} implies that $
{\bm Y} = ({\bm I} - \overline{\bm D})^{-1} \overline{\bm B} {\bm Z}$. Fig.~\ref{fig:col} illustrates the 
relationship between the data matrices/vectors used in the social network sensing problem. 
One often obtains a noisy estimate of ${\bm Y}$; i.e.,
\begin{equation} \label{eq:constraint}
\hat{\bm Y} = ({\bm I} - \overline{\bm D})^{-1} \overline{\bm B} {\bm Z} + {\bm N}.
\end{equation}
We relegate the discussion on obtaining $\hat{\bm Y}$ when the agents' interactions are asynchronous to Section~\ref{sec:random}.

From Eq.~\eqref{eq:constraint}, a natural solution to estimate $(\overline{\bm B}, \overline{\bm D})$ is to obtain a tuple $({\bm B}, {\bm D})$ that minimizes the square loss $\| \hat{\bm Y} - ({\bm I} - {\bm D})^{-1} {\bm B} {\bm Z}  \|_F^2$. However, the system equation \eqref{eq:constraint} admits an implicit ambiguity:
\begin{Lemma} \label{lem:amb}
Consider the pair of tuples, $({\bm B}, {\bm D})$ and $(\tilde{\bm B}, \tilde{\bm D})$, where ${\bm B}{\bm 1} + {\bm D} {\bm 1} = {\bm 1}$ and the latter is defined as
\begin{subequations} \label{eq:equiv}
\begin{align}
\tilde{\bm B} = \bm{\Lambda} {\bm B},~
{\rm off}(\tilde{\bm D}) = \bm{\Lambda} {\rm off}({\bm D}), \label{eq:amb_a} \\
{\rm diag}(\tilde{\bm D}) = {\bf 1} - \bm{\Lambda} ( {\bm B} {\bf 1} + {\rm off}({\bm D}) {\bf 1} ), \label{eq:amb_b}
\end{align}
\end{subequations}
for some diagonal matrix $\bm{\Lambda} > {\bm 0}$ such that ${\rm diag}(\tilde{\bm D}) \geq {\bm 0}$.
Under \eqref{eq:equiv}, the equalities $({\bm I} - {\bm D})^{-1} {\bm B} = ({\bm I} - \tilde{\bm D})^{-1} \tilde{\bm B}$ and $\tilde{\bm B} {\bm 1} + \tilde{\bm D} {\bm 1 } = {\bm 1}$ hold.
\end{Lemma}
The proof is relegated to Appendix~\ref{app:lemamb}. Lemma~\ref{lem:amb} indicates that there are infinitely many tuples $(\tilde{\bm B}, \tilde{\bm D})$ with the same product $({\bm I} - \tilde{\bm D})^{-1} \tilde{\bm B}$. The diagonal entries of $\tilde{\bm D}$; i.e., the magnitude of \emph{self trust}, are ambiguous. In fact, the ambiguity described in Lemma~\ref{lem:amb} is due to the fact that the collected data $\hat{\bm Y}, {\bm Z}$ lacks information about the rate of social interaction.

In light of Lemma~\ref{lem:amb}, we determine an \emph{invariant} quantity among the ambiguous solutions.
Define the equivalence class:
\begin{equation}
({\bm B}, {\bm D}) \sim (\tilde{\bm B}, \tilde{\bm D}) : \exists~ \bm{\Lambda} > {\bm 0}~{\rm s.t.}~\eqref{eq:equiv}~{\rm holds},~{\rm diag}(\tilde{\bm D}) \geq {\bm 0}.
\end{equation}
A quick observation on \eqref{eq:equiv} is that $\bm{\Lambda}$ modifies the magnitude of ${\rm diag}({\bm D})$.
This inspires us to resolve the ambiguity issue by imposing constraints on ${\rm diag}({\bm D})$:

\begin{Observation} \label{obs:r}
The pair of {relative trust matrices} resulting from $({\bm B}, {\bm D})$, denoted by the superscript $(\cdot)'$:
\begin{equation}
\begin{array}{l}
{\bm B}' = ({\bm I} - {\rm Diag}({\bm c})) \bm{\Lambda}_s^{-1} {\bm B},~{\rm diag}({\bm D}') = {\bm c}, \\
{\rm off}({\bm D}' ) = ({\bm I} - {\rm Diag}({\bm c}))  \bm{\Lambda}_s^{-1} {\rm off}({\bm D}) ,
\end{array}
\end{equation}
where $\bm{\Lambda}_s = {\bm I} - {\rm Diag}( {\rm diag}( {\bm D} ))$ and ${\bm 0} \leq {\bm c} < {\bm 1}$, is {unique} for each equivalence class when ${\bm c}$ is fixed.  In other words, if $({\bm B}, {\bm D}) \sim (\tilde{\bm B}, \tilde{\bm D})$, their resultant pairs of relative trust matrices are equal, $({\bm B}', {\bm D}') = (\tilde{\bm B}', \tilde{\bm D}')$.
\end{Observation}
Notice that the pair of relative trust matrices is an important quantity for a social network since it contains the relative trust strengths and connectivity of the network.

We are now ready to present the regression problems pertaining to recovering $\tilde{\bm B}', \tilde{\bm D}'$. As we often have access to a superset ${\cal S}$ of the support of $\overline{\bm D}$; i.e., $\Omega_{\overline{\bm D}} \subseteq {\cal S}$ and $\Omega_{\overline{\bm D}}$ denotes the support of $\overline{\bm D}$, we propose two different formulations when different knowledge on $\Omega_{\overline{\bm D}}$ is accessible.

\emph{\textbf{Case 1: nearly full support information ---}} When $\Omega_{\overline{\bm D}} \subseteq {\cal S}$ and $|{\cal S}| \approx |\Omega_{\overline{\bm D}}|$, the support of $\overline{\bm D}$ is mostly known. We  formulate the following least square problem:
\begin{subequations} \label{eq:nsi_full}
\begin{align}
\displaystyle \min_{ {\bm D}, {\bm B} }~ & \| ({\bm I} - {\bm D}) \hat{\bm Y} - {\bm B} {\bm Z} \|_F^2 \\
{\rm s.t.}~ & {\bm B}{\bf 1} + {\bm D} {\bf 1} = {\bf 1}, ~{\bm D} \geq {\bf 0},~ {\bm B} \geq {\bm 0},\label{eq:const_full1} \\
~& {\cal P}_{\Omega_{\overline{\bm B}}^c} ({\bm B})= {\bm 0}, {\cal P}_{{\cal S}^c} ({\bm D})= {\bm 0}, {\rm diag}({\bm D}) = {\bm c} \label{eq:const_full2}.
\end{align}
\end{subequations}
The projection operator ${\cal P}_{\Omega} (\cdot)$ is defined as:
\begin{equation}
[{\cal P}_{\Omega} ( {\bm A} )]_{ij} = \begin{cases}
A_{ij},~ij \in \Omega,~\\
0,~{\rm otherwise,}
\end{cases}
\end{equation}
where $\Omega \subseteq [n] \times [m]$ is an index set for the matrix ${\bm A} \in \mathbb{R}^{n \times m}$.
Problem~\eqref{eq:nsi_full} is a constrained least-square problem that can be solved efficiently,  e.g., using \texttt{cvx} \cite{cvx}.

\emph{\textbf{Case 2: partial support information ---}}  When $\Omega_{\overline{\bm D}} \subseteq {\cal S}$ and $|{\cal S}| \gg |\Omega_{\overline{\bm D}}|$, the system equation \eqref{eq:constraint} constitutes an undetermined system. This motivates us to exploit that $\overline{\bm D}$ is sparse and consider the sparse recovery problem:
\begin{subequations} \label{eq:nsi_first2}
\begin{align}
\displaystyle \min_{ {\bm D}, {\bm B} }~ & \| {\rm vec} (  {\bm D} ) \|_0 \\
{\rm s.t.}~ & \| ({\bm I} - {\bm D}) \hat{\bm Y} - {\bm B} {\bm Z} \|_F^2 \leq \epsilon,
~{\bm B}{\bf 1} + {\bm D} {\bf 1} = {\bf 1},
\label{eq:const_a}  \\
~& {\bm D} \geq {\bf 0},~{\bm B} \geq {\bm 0},~{\rm diag}({\bm D}) = {\bm c}, \label{eq:const_c} \\
~& {\cal P}_{\Omega_{\overline{\bm B}}^c} ({\bm B})= {\bm 0},~{\cal P}_{{\cal S}^c} ({\bm D})= {\bm 0}, \label{eq:const_b}
\end{align}
\end{subequations}
for some $\epsilon \geq 0$ and {  ${\rm vec}({\bm D})$ denotes the vectorization of the matrix ${\bm D}$}. Note that Problem~\eqref{eq:nsi_first2} is an $\ell_0$ minimization problem that is NP-hard to solve in general. In practice, the following convex  problem is solved in lieu of \eqref{eq:nsi_first2}:
\begin{equation} \label{eq:nsi_l1}
\min_{ {\bm B}, {\bm D}} \| {\rm vec} ({\bm D} ) \|_1~{\rm s.t.}~{\rm Eq.}~\eqref{eq:const_a}-\eqref{eq:const_b}.
\end{equation}

\begin{Remark}
A social network sensing problem similar to Case 1 has been studied in \cite{Khan2010} and its identifiability conditions have been discussed therein. In fact, we emphasize that our focus in this paper is on the study of Case 2, which presents a  more challenging problem due to the lack of support information.
\end{Remark}

\subsection{Identifiability Conditions for Social Network Sensing}
We derive the conditions for recovering the relative trust matrix $(\overline{\bm B}', \overline{\bm D}')$ resulting from $(\overline{\bm B}, \overline{\bm D})$ by solving \eqref{eq:nsi_full} or \eqref{eq:nsi_first2}.
As a common practice in signal processing, the following analysis assumes noiseless measurements such that $\sigma^2 = {\bm 0}$. We set $\epsilon = 0$ in \eqref{eq:const_a} and assume $K \geq n_s$. 

As $\sigma^2 = 0$, the optimization problem \eqref{eq:nsi_full} admits an optimal objective value of zero. Now, let us denote
\begin{equation} \label{eq:a_tilde}
\tilde{\bm A} \triangleq
\left( \begin{array}{cc} \hat{\bm Y}^T & {\bm Z}^T \end{array} \right).
\end{equation}
The identifiability condition for \eqref{eq:nsi_full} and \eqref{eq:nsi_first2} can be analyzed by studying the linear system: 
\begin{equation}\label{eq:equal_full}
\hat{\bm y}_i = \tilde{\bm A} \left( \begin{array}{c} {\bm d}_i \\ {\bm b}_i \end{array} \right),~\forall~i
,~{\rm Eq.}~\eqref{eq:const_full1} -\eqref{eq:const_full2},
\end{equation}
where $\hat{\bm y}_i$ is the $i$th column of $\hat{\bm Y}$, ${\bm d}_i, {\bm b}_i$ are the $i$th row of ${\bm D}, {\bm B}$, respectively.
Note that the above condition is equivalent to \eqref{eq:const_a}--\eqref{eq:const_b} with $\epsilon = 0$. 
{  In the following, we denote ${\cal S}_i$ and $\Omega_{\overline{\bm B}}^i$ as the index sets restricted to the $i$th row of ${\bm D}$ and ${\bm B}$}. 

\emph{\textbf{Case 1: nearly full support information ---}} 
Due to the constraints in \eqref{eq:const_full2}, the number of unknowns in the right hand side of the first equality in \eqref{eq:equal_full} is $|\Omega_{\overline{\bm B}}^i| + |{\cal S}_i| - 1$. From linear algebra, the following is straightforward to show:
\begin{Prop}\label{prop:full}
The system \eqref{eq:equal_full} admits a unique solution if
\begin{equation} \label{eq:fullsupport}
{\rm rank} ( \tilde{\bm A}_{:, {\cal S}_i\cup\Omega_{\overline{\bm B}}^i} ) \geq |\Omega_{\overline{\bm B}}^i| + |{\cal S}_i| - 1,~\forall~i,
\end{equation}
where $\tilde{\bm A}_{:, {\cal S}_i \cup\Omega_{\overline{\bm B}}^i}$ is a submatrix of $\tilde{\bm A}$ with the columns taken from the respective entries of ${\cal S}_i$ and $\Omega_{\overline{\bm B}}^i$.
\end{Prop}
{  Similar result has also been reported in \cite[Lemma 10]{Khan2010}.}

\emph{\textbf{Case 2: partial support information ---}}
As $|{\cal S}| \gg |\Omega_{\overline{\bm D}}|$, this case presents a more challenging scenario to analyze since \eqref{eq:equal_full} is an underdetermined system.
In particular, \eqref{eq:fullsupport} is often not satisfied. 
However, a sufficient condition can be given by exploiting the fact that \eqref{eq:nsi_first2} finds the sparsest solution:
\begin{Prop}\label{prop:partial}
There is a unique optimal solution to \eqref{eq:nsi_first2} if for all $\tilde{\cal S}_i \subseteq {\cal S}_i$ and $|\tilde{\cal S}_i| \leq 2 | \Omega_{\overline{\bm D}}^i|$, we have
\begin{equation} \label{eq:partialsupport}
{\rm rank} ( \tilde{\bm A}_{:, \tilde{\cal S}_i\cup\Omega_{\overline{\bm B}}^i} ) \geq |\Omega_{\overline{\bm B}}^i| + 2|\Omega_{\overline{\bm D}}^i| - 1,~\forall~i,
\end{equation}
where $\tilde{\bm A}_{:, \tilde{\cal S} \cup\Omega_{\overline{\bm B}}^i}$ is a submatrix of $\tilde{\bm A}$ with the columns taken from the respective entries of ${\cal S}_i$ and $\Omega_{\overline{\bm B}}^i$.
\end{Prop}
Proposition~\ref{prop:partial} is equivalent to characterizing the \emph{spark} of the matrix $\tilde{\bm A}$; see \cite{Eldar2014}. 

Checking \eqref{eq:partialsupport} is non-trivial and it is not clear how many stubborn agents are needed. We next show that using an optimized placement of stubborn agents, one can derive a sufficient condition for unique recovery using \eqref{eq:nsi_first2}.

\subsection{Optimized placement of stubborn agents} 
We consider an \emph{optimized placement of stubborn agents} when only partial support information is given. 
In other words, we design $\Omega_{\overline{\bm B}}$ for achieving better sensing performance.  
This is possible in a controlled experiment where the stubborn agents are directly controlled to 
interact with a target group of ordinary agents. 
Note also that \cite{yildiz2013binary} has studied an optimal stubborn agent placement 
formulation for the voter's model opinion dynamics with a different objective. 

To fix ideas, the following discussion draws a connection between 
\emph{blind compressed sensing} and our social network sensing problem. 
Consider the following equivalent form of \eqref{eq:const_a}:
\begin{equation}
({\bm Y} {\bm Z}^\dagger )^T ={\bm B}^{T}  ({\bm I} - {\bm D})^{-T},
\end{equation}
where ${\bm Z}^\dagger$ denotes the psuedo-inverse of ${\bm Z}$. By noting that ${\bm Y}{\bm Z}^\dagger = ({\bm I} - \overline{\bm D}')^{-1} \overline{\bm B}'$, we have:
\begin{equation} \label{eq:linear_rel}
\begin{array}{c}
\overline{\bm B}^{'T} ({\bm I} - \overline{\bm D}')^{-T} = {\bm B}^{T}  ({\bm I} - {\bm D})^{-T} \\
\Longleftrightarrow \overline{\bm B}^{'T} ({\bm I} - \overline{\bm D}')^{-T}  (\overline{\bm D}' - {\bm D} )^T = ({\bm B} - \overline{\bm B}')^{T} \\
\Longleftrightarrow \overline{\bm B}^{'T} ({\bm I} - \overline{\bm D}')^{-T}  (\overline{\bm d}_i' - {\bm d}_i ) = {\bm b}_i - \overline{\bm b}_i',~\forall~i,
\end{array}
\end{equation}
where $\overline{\bm d}_i', {\bm d}_i, \overline{\bm b}_i'$ and ${\bm b}_i$ are the $i$th row of $\overline{\bm D}', {\bm D}, \overline{\bm B}'$ and ${\bm B}$, respectively.
Due to the constraint ${\cal P}_{{\cal S}^c} ({\bm D})= {\bm 0}$ and ${\rm diag}({\bm D}) = {\bm c}$, the number of unknowns in ${\bm d}_i$ is $n_i \triangleq n - n_s - s_i - 1$, where $s_i = |{\cal S}^c_i|$ and ${\cal S}^c_i$ is the complement of support set restricted to the $i$th row of ${\bm D}$.

The matrix $\overline{\bm B}^{'T} ({\bm I} - \overline{\bm D}')^{-T}$ can be regarded as a sensing matrix in \eqref{eq:linear_rel}.
Note that because $\overline{\bm B}'$  is  unknown, we have a \emph{blind compressed sensing} problem, which was studied in \cite{Gleichman2011,Studer2012}.
However, the scenarios considered there were different from ours since  the zero values of the sensing matrix are partially known in that case.

To study an identifiability condition for \eqref{eq:linear_rel}, we note that $\overline{\bm B}^{'T} ({\bm I} - \overline{\bm D}')^{-T} = \overline{\bm B}^{'T}  ({\bm I} + \overline{\bm D}' + (\overline{\bm D}')^2 + \ldots)^T$; i.e., the sensing matrix is equal to a perturbed $\overline{\bm B}^{'T}$. When the perturbation is small, we could study $\overline{\bm B}$ alone.
Moreover, as the entries in $\overline{\bm B}$ are unknown, it is desirable to consider a sparse construction to reduce the number of unknowns.

As suggested in \cite{Khajehnejad2011,Wang2011}, a good choice is to construct $\Omega_{\overline{\bm B}}$ randomly such that each row in $\overline{\bm B}$ has a constant number $d$ of non-zero elements (independent of $n_i'$). 
We have the following sufficient condition:
\begin{Theorem} \label{thm:cs}
{  Define ${\rm supp}(\overline{\bm B}) = \{ ij : \overline{B}_{ij} > 0 \}$, $b_{min} = \min_{ij \in {\rm supp}(\overline{\bm B})} \overline{B}_{ij}'$, $b_{max} = \max_{ij \in {\rm supp}(\overline{\bm B})} \overline{B}_{ij}'$, $ \beta_i = n_s / n_i$ and $\beta_i' = \beta_i - d/n_i$. }
Let the support of $\overline{\bm B}' \in \mathbb{R}^{n \times n_s}$ be constructed such that each row  has $d$ non-zero elements, selected randomly and independently. If
\begin{equation} \label{eq:thmcs}
d > \max \Big\{ 4, 1 + \frac{H(\alpha_i) + \beta' H(\alpha_i/\beta_i')}{\alpha_i \log (\beta_i' / \alpha_i)} \Big\},
\end{equation}
and
\begin{equation} \label{eq:thmval}
b_{min} (2 d - 3) - 1 - 2 b_{max} > 0,
\end{equation}
where $H(x)$ is the binary entropy function,
and $ \| \overline{\bm d}_i' \|_0 \leq \alpha_i n_i / 2$ for all $i$,  then as $n-n_s \rightarrow \infty$, there is a unique optimal solution to \eqref{eq:nsi_first2} that $({\bm B}^\star, {\bm D}^\star) = (\overline{\bm B}', \overline{\bm D}')$ with probability one.
Moreover, the failure probability is bounded as:
\begin{equation}
\begin{array}{l}
\displaystyle {\rm Pr}( ({\bm B}^\star, {\bm D}^\star) \neq (\overline{\bm B}', \overline{\bm D}') ) \vspace{.2cm} \\
\displaystyle  ~~~~\leq \max_i \left( \frac{d}{\beta_i} \right)^4 \frac{d-1}{{n_i}^2} + {\cal O}( {n_i}^{2- (d-1)(d-3)} ).
 \end{array}
\end{equation}
\end{Theorem}
The proof of Theorem~\ref{thm:cs} is in Appendix~\ref{app:thmcs} where the claim is proven by treating the unknown entries of $\overline{\bm B}^{'T}$ as \emph{erasure bits}, and showing that the sensing matrix with erasure corresponds to a high quality expander graph with high probability. To the best of our knowledge, Theorem~\ref{thm:cs} is a new recoverability result proven for blind compressed sensing problems.

Condition~\eqref{eq:thmcs} gives a guideline for choosing the number of stubborn agents  $n_s$ employed. In fact, if we set $\alpha \rightarrow 0$ while keeping the ratio $\beta / \alpha$ constant, condition \eqref{eq:thmcs} can be approximated by $\beta > \alpha = 2 p$, where  $\| \overline{\bm d}_i' \|_0 \leq p(n-n_s)$. {  Notice that in the limit as $n - n_s \rightarrow \infty$, if every ordinary agent in the sub-network that corresponds to $\overline{\bm D}$ has an in-degree bounded by $p(n-n_s)$, then we only need:
\begin{equation}
n_s \geq 2 p (n-n_s)
\end{equation}
stubborn agents to perfectly reconstruct $\overline{\bm D}$.} 

On the other hand, condition \eqref{eq:thmval} indicates that the amount of \emph{relative trust} in stubborn agents cannot be too small. This is reasonable in that the network sensing performance should depend on the influencing power of the stubborn agents.
The effect of the known support is also reflected. In particular, an increase in $s_i$ leads to a decrease in $n_i$ and an increase in $\beta_i$. The maximum allowed sparsity $\alpha_i$ is increased as a result.
We conclude this subsection with the following remarks.
\begin{Remark}
When $n$ is finite, the failure probability may grow with the size of $\Omega_{\overline{\bm B}}$, i.e., it is ${\cal O}(d^5)$, coinciding with the intuition concerning the tradeoff between the size of $\Omega_{\overline{\bm B}}$  and the sparse recovery performance. Although the failure probability vanishes as $n - n_s \rightarrow \infty$, the parameter $d$ should be chosen judiciously in the design.
\end{Remark}
\begin{Remark} \label{rem:sparse}
Another important issue that affects the recoverability of \eqref{eq:nsi_first2} is the degree distribution in $G$ {  as the conditions in Theorem~\ref{thm:cs} requires the sparsity level of $\overline{\bm D}$ to be small for every row.} Given a fixed total number of  edges, it is easier to recover a network with a concentrated degree distribution (e.g., the Watts-Strogatz network \cite{Watts1998}) while a network with power law degree distribution (e.g., the Barabasi-Albert network \cite{Albert2002}) is more difficult to recover.
\end{Remark}

\begin{Remark}{ 
The requirement on the number of stubborn agents in Theorem~\ref{thm:cs} may appear to be restrictive. However, note that only the expected value $\overline{\bm B} $ is considered in the model. 
Theoretically, one only need to employ a small number of stubborn agents that \emph{randomly} wander in different positions of the social network and `act' as agents with different opinion,  thus achieving an effect similar to that of a {\it synthetic aperture RADAR}. 
This effectively creates a vast number of \emph{virtual} stubborn agents from a small number of \emph{physically present}  stubborn agents.}
\end{Remark}

\section{Implementations of Social Network Sensing} \label{sec:fast}
In this section we discuss practical issues involved in implementing our network sensing method. First, we develop a fast algorithm for solving large-scale network sensing problems. Second, we consider a random opinion dynamics model and propose a consistent estimator for the steady state.

\subsection{Proximal Gradient Method for Network Sensing} \label{sec:fast2}
This subsection presents a practical implementation method for large-scale network sensing when $n \gg 0$. We resort to a first order optimization method.  The main idea is to employ a proximal gradient method \cite{Parikh2014} to the following problem:
\begin{equation} \label{eq:lag_rel}
\begin{array}{rl}
\displaystyle \min_{ {\bm D}, {\bm B} } & \lambda \| {\rm vec} ( {\bm D} ) \|_1 + f({\bm B}, {\bm D}) \\
{\rm s.t.} & {\bm B} \geq {\bf 0},~ {\bm D} \geq {\bf 0},~{\cal P}_{{\cal S}^c} ( {\bm D} ) = {\bm 0}, \\
& {\cal P}_{\Omega_{\overline{\bm B}}^c} ({\bm B}) = {\bm 0},~{\rm diag}({\bm D}) = {\bm c},
\end{array}
\end{equation}
 where
\[
f({\bm B}, {\bm D}) =  \| ({\bm I} - {\bm D})\hat{\bm Y}{\bm Z}^{\dagger} - {\bm B} \|_F^2 + \gamma \| {\bm B}{\bf 1} + {\bm D} {\bf 1} - {\bf 1} \|_2^2
\]
and $\gamma > 0$.
Note that ${\bm Z}^{\dagger}$ is the right pseudo-inverse of ${\bm Z}$.
Problem \eqref{eq:lag_rel} can be seen as the Lagrangian relaxation
of \eqref{eq:nsi_l1}. Meanwhile, similar to the first case considered in Section~\ref{sec:netwk}, we take $\lambda = 0$ when $|{\cal S}| \approx |\Omega_{\overline{\bm D}}|$.

The last term in \eqref{eq:lag_rel} is continuously differentiable. Let us denote the feasible set of \eqref{eq:lag_rel} as ${\cal L}$:
\begin{equation}
\begin{split}
{\cal L} = & \{ ({\bm B}, {\bm D}) : {\bm B} \geq {\bf 0},~{\bm D} \geq {\bf 0},~{\cal P}_{{\cal S}^c} ( {\bm D} ) = {\bm 0}  \\
& {\cal P}_{\Omega_{\overline{\bm B}}^c} ({\bm B}) = {\bm 0},~{\rm diag}({\bm D}) = {\bm c} \}.
\end{split}
\end{equation}
Let $\ell \in \mathbb{N}$ be the iteration index of the proximal gradient method. At the $\ell$th iteration, we  perform the following update:
\begin{equation} \label{eq:upd}
\begin{array}{rl} \hspace{-.4cm}
\displaystyle ( {\bm B}^{\ell}, {\bm D}^\ell ) =  \arg  \hspace{-.2cm} \min_{({\bm B}, {\bm D}) \in {\cal L} } \hspace{-.5cm} &  \hspace{.4cm} \alpha \lambda \| {\rm vec} ({\bm D}) \|_1  \\
& \hspace{-.5cm}+ \| {\bm B} - \tilde{\bm B}^{\ell-1} \|_F^2 + \| {\bm D} - \tilde{\bm D}^{\ell-1} \|_F^2
\end{array}\hspace{-.2cm}
\end{equation}
where
\begin{subequations} \label{eq:gradient}
\begin{align}
\tilde{\bm D}^{\ell} & = {\bm D}^{\ell} - \alpha \nabla_{{\bm D}} f({\bm B}^\ell, {\bm D}) \big|_{{\bm D} = {\bm D}^\ell} \\
\tilde{\bm B}^{\ell} & = {\bm B}^{\ell} - \alpha \nabla_{{\bm B}} f({\bm B}, {\bm D}^\ell) \big|_{{\bm B} = {\bm B}^\ell},
\end{align}
\end{subequations}
and $\alpha > 0$ is a fixed step size such that $\alpha < 1/L$, where $L$ is the Lipschitz constant for the gradient of $f$.

Importantly, the proximal update \eqref{eq:upd} admits a closed form solution using the soft-thresholding operator:
\begin{subequations}
\begin{align}
{\bm B}^\ell & = \max \{ {\bm 0}, {\cal P}_{\Omega_{\overline{\bm B}}^c} (\tilde{\bm B}^{\ell-1}) \} \\
{\rm off}({\bm D}^\ell) & = {\cal P}_{{\cal S}^c} (  \texttt{soft\_th}_{\alpha \lambda} ( {\rm off} (\tilde{\bm D}^{\ell-1} )) ),
\end{align}
\end{subequations}
where $\texttt{soft\_th}_{\lambda} (\cdot)$ is a \emph{one-sided} soft thresholding operator \cite{Beck2009} that applies element-wisely and $\texttt{soft\_th}_{\lambda} (x) = u(x) \max\{0, x-\lambda\} $.
To further accelerate the algorithm, we adopt an update rule similar to the fast iterative shrinkage (FISTA) algorithm in \cite{Beck2009}, as summarized in Algorithm~\ref{alg:px}. As seen, the per-iteration complexity is ${\cal O}( n^2 + n\cdot n_s)$; i.e., it is linear in the number of variables.

\algsetup{indent=1em}
\begin{algorithm}[t]
\caption{FISTA algorithm for \eqref{eq:lag_rel}.}\label{alg:px}
  \begin{algorithmic}[1]
  \STATE \textbf{Initialize:} $({\bm B}^0, {\bm D}^0) \in {\cal L}$, $t_0 = 0$, $\ell =1$;
  \WHILE {\emph{convergence is not reached}}
   \STATE Compute the proximal gradient update direction:
\begin{subequations}
\begin{align}
\bm{dB}^\ell & \leftarrow \max \{ {\bm 0}, {\cal P}_{\Omega_{\overline{\bm B}}^c} (\tilde{\bm B}^{\ell-1}) \} \nonumber \\
\bm{dD}^\ell & \leftarrow {\cal P}_{{\cal S}^c} ( \texttt{soft\_th}_{\alpha \lambda} ({\rm off}(\tilde{\bm D}^{\ell-1})) ), \nonumber
\end{align}
\end{subequations}
   where $\tilde{\bm B}^{\ell-1}, \tilde{\bm D}^{\ell-1}$ are given by \eqref{eq:gradient}.
   \STATE Update $t_{\ell} \leftarrow (1 + \sqrt{1 + 4 t_{\ell-1}^2}) / 2$.
   \STATE Update the variables as:
   \begin{subequations}
\begin{align}
\bm{B}^k & \leftarrow \bm{dB}^\ell + \frac{t_{\ell-1} - 1}{t_\ell} ( \bm{dB}^\ell - \bm{dB}^{\ell-1} ) \nonumber \\
{\rm off}(\bm{D}^\ell) & \leftarrow \bm{dD}^\ell + \frac{t_{\ell-1} - 1}{t_\ell} ( \bm{dD}^\ell - \bm{dD}^{\ell-1} ) \nonumber , \nonumber
\end{align}
\end{subequations}
   \STATE $\ell \leftarrow \ell+1.$
\ENDWHILE
\STATE Set ${\rm diag}({\bm D}^\ell) \leftarrow {\bm c}$.
\STATE \textbf{Return:} $({\bm B}^\ell, {\bm D}^\ell)$.
  \end{algorithmic}
\end{algorithm}

We conclude this subsection with a discussion on the convergence of Algorithm~\ref{alg:px}. As \eqref{eq:lag_rel} is convex and $f$ is continuously differentiable, the proximal gradient method using \eqref{eq:upd} is guaranteed \cite{Beck2009} to converge to an optimal solution of \eqref{eq:lag_rel}. Moreover, the convergence speed is ${\cal O}(1/\ell^2)$.

\subsection{Random Opinion Dynamics} \label{sec:random}
So far, our method for network sensing only requires collecting the asymptotic states $\mathbb{E}\{{\bm y}(\infty;k) | {\bm x}(0;k) \}$.
Importantly, the data matrices $\hat{\bm Y}$ and ${\bm Z}$ (cf.~\eqref{eq:collect}) can be collected easily when the trust matrix ${\bm W}(t;k)$ is deterministic; i.e., ${\bm W}(t;k) = \overline{\bm W}$ for all $t,k$, and the observations are noiseless. For the case where ${\bm D}(t;k)$ is random, computing the expectation may be difficult since the latter is an average taken over an ensemble of the sample paths of $\{ {\bm W}(t;k) \}_{\forall t, \forall k}$. 

{  This section considers adopting the proposed active sensing method to the case with randomized opinion dynamics, which captures the possible randomness in social interactions.} 
Our idea is to propose a consistent estimator for $\mathbb{E}\{{\bm y}(\infty;k) | {\bm z}(0;k) \}$ using the opinions gathered from the same time series $\{ \hat{\bm x}(t;k) \}_{t \in {\cal T}_k}$, where ${\cal T}_k \subseteq \mathbb{Z}^+$ is now an arbitrary sampling set. Specifically, we show that the random process ${\bm x}(t;k)$ is ergodic.
We first make an interesting observation pertaining to random opinion dynamics:
\begin{Observation} \label{obs:fluctuate} \cite[Theorem 2]{Acemoglu2013} \cite{bianchi2012}
When $n_s \geq 2$ and under a random opinion dynamics model, the opinions will not converge; i.e., ${\bm x}(t;k) \neq {\bm x}(t-1;k)$ almost surely.\vspace{.1cm}
\end{Observation}
Observation~\ref{obs:fluctuate} suggests that a natural approach to computing the expectation $\mathbb{E}\{{\bm y}(\infty;k) | {\bm x}(0;k) \}$ is by taking averages over the temporal samples. We propose the following estimator:
\begin{equation} \label{eq:sample}
\mathbb{E}\{{\bm x}(\infty;k) | {\bm x}(0;k) \} \approx \hat{\bm x}({\cal T}_k; k) \triangleq \frac{1}{ |{\cal T}_k|} \sum_{t \in {\cal T}_k} \hat{\bm x}(t;k),
\end{equation}
where we recall the definition for $\hat{\bm x}(t;k)$ from \eqref{eq:obs}.
Notice that $\mathbb{E}\{{\bm y}(\infty;k) | {\bm x}(0;k) \}$ can be retrieved from $\mathbb{E}\{{\bm x}(\infty;k) | {\bm x}(0;k) \}$ by taking the last $n-n_s$ elements of the latter.
In order to compute the right hand side in \eqref{eq:sample}, we only need to know the cardinality of ${\cal T}_k$; i.e., the number of samples collected. Knowledge on the members in ${\cal T}_k$ is not required.
Specifically, the temporal samples can be collected through random (and possibly noisy) sampling at time instances on the opinions.
The following theorem characterizes the performance of \eqref{eq:sample}:
\begin{Theorem} \label{thm:asymp}
Consider the estimator in \eqref{eq:sample} with a sampling set ${\cal T}_k$. Denote $\overline{\bm x}(\infty; k) \triangleq \lim_{t \rightarrow \infty} \mathbb{E} \{ {\bm x} (t;k) | {\bm x}(0;k) \} = \overline{\bm W}^\infty {\bm x}(0;k)$ and assume that $\mathbb{E} \{ \| {\bm D}(t;k) \|_2 \} < 1$. If $T_o \rightarrow \infty$, then
\begin{enumerate}
\item the estimator \eqref{eq:sample} is unbiased:
\begin{equation}
\mathbb{E} \{ \hat{\bm x}({\cal T}_k; k) | {\bm z}(0;k)\} =  \overline{\bm x}(\infty; k).
\end{equation}
\item
 the estimator \eqref{eq:sample} is asymptotically consistent:
\begin{equation} \label{eq:est_cons}
\lim_{ |{\cal T}_k| \rightarrow \infty } \mathbb{E} \{ \| \hat{\bm x}({\cal T}_k; k) - \overline{\bm x}(\infty;k) \|_2^2 | {\bm x}(0;k) \} = 0.\vspace{.2cm}
\end{equation}
\end{enumerate}
For the latter case, we have
\begin{equation} \label{eq:final_bd}
\begin{array}{l}
\displaystyle  \mathbb{E} \{ \| \hat{\bm x}({\cal T}_k;k) - \overline{\bm x}(\infty;k) \|_2^2 | {\bm x}(0;k) \} \\
\displaystyle ~~\leq \frac{{C'}}{|{\cal T}_k|} \Big( \sum_{i=0}^{|{\cal T}_k|-1} \lambda^{\min_{\ell} |t_{\ell+i} - t_\ell|}  \Big),
\end{array}
\end{equation}
where $C' < \infty$ is a constant and $\lambda = \lambda_{max} (\overline{\bm D}) < 1$, i.e., the latter term is a geometric series with bounded sum.
\end{Theorem}
Note that a similar result to Theorem~\ref{thm:asymp} was reported in \cite{Ravazzi2015}. Our result is specific to the case with stubborn agents, which allows us to find a precise characterization of the mean square error. The proof of Theorem~\ref{thm:asymp} can be found in Appendix~\ref{app:cons}.

\begin{Remark}
From \eqref{eq:final_bd}, we observe that the upper bound on the mean square error can be optimized by maximizing $\min_{i,j, i \neq j} |t_i - t_j|$.
Suppose that the samples $\hat{\bm x}({\cal T}_k;k)$ have to be taken from a finite  interval $[T_{max}] \setminus [T_o]$, $T_{max} < \infty$ and $|{\cal T}_k| < \infty$; here, the best estimate can be obtained by using sampling instances that are drawn uniformly from $[T_{max}] \setminus [T_o]$.
\end{Remark}

\section{Numerical Results} \label{sec:sim}
To validate our methods, we conducted several numerical simulations, reconstructing both synthetic networks and real networks. In this section, we focus on cases where the network dynamics model \eqref{eq:dyn_mat} is exact (while the measurement may be noisy),but emphasize the crucial importance of considering data collected from real networks, e.g., the online social networks (e.g., Facebook, Twitter).
The Monte-Carlo simulations were 
obtained by averaging over at least $100$ simulation trials. {  We also set $K = 2 n_s$ and ${\bm c} = {\bm 0}$ to respect the requirement $K \geq n_s$ and for ease of comparison.}

\subsection{Synthetic networks with noiseless measurement}
We evaluate the sensing performance on a synthetic network with noiseless measurement on the steady system state.
In light of Theorem~\ref{thm:cs}, for the placement of the stubborn agents, we randomly connect $d$ stubborn agents to each ordinary agent.

\begin{figure}[t]
\centering
\includegraphics[width =0.49\linewidth]{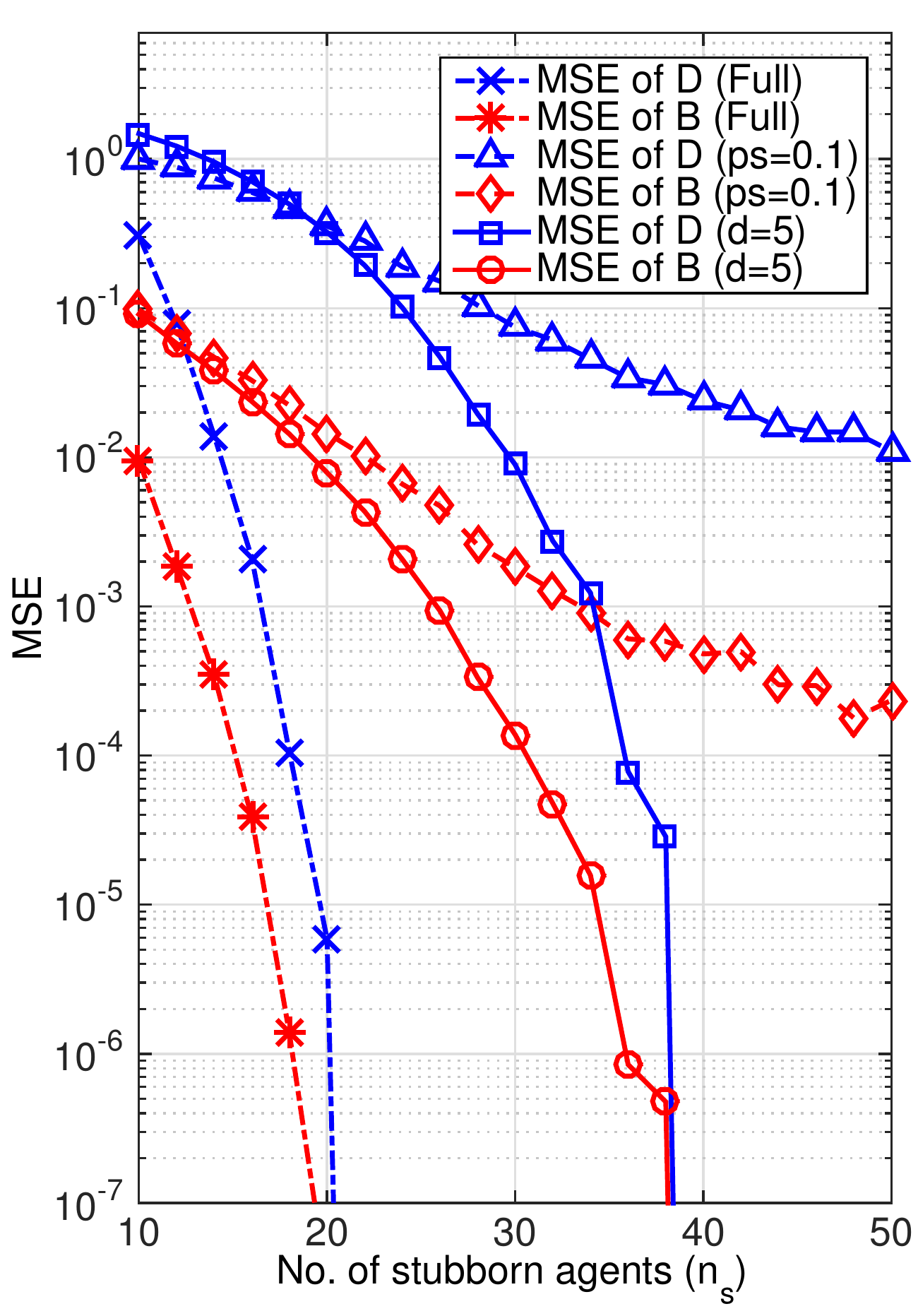} \includegraphics[width =0.475\linewidth]{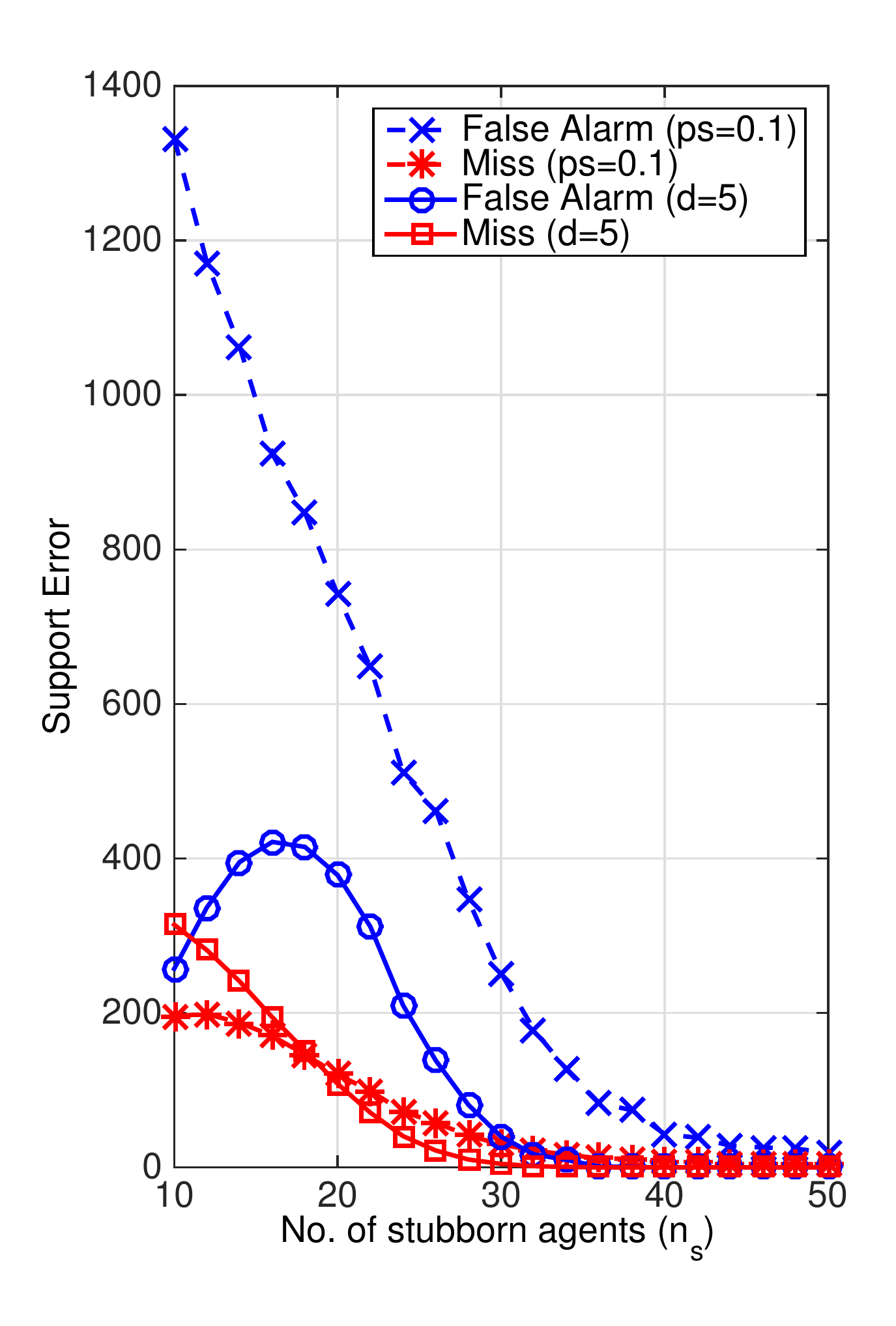}
\caption{Comparing performance against the number of stubborn agents $n_s$. (Left) The NMSE. (Right) The average support recovery error. In the legend, `full' denotes the case with full support information; `(ps=0.1)' and `(d=5)' denotes the case where $\overline{\bm B}$ is constructed as a random bipartite graph and a random $d$-regular bipartite graph, respectively.} \vspace{-.2cm}
\label{fig:mse_vs_ns}
\end{figure}

The first numerical example compares performance in recovering $\overline{\bm D}'$ and $\overline{\bm B}'$ against the number of stubborn agents $n_s$. We fix the number of ordinary agents at $n - n_s=60$. The network $G$ is constructed as an Erdos-Renyi (ER) graph with connectivity $p=0.1$. Furthermore, the weights in $\overline{\bm W}$ are generated uniformly first, and normalized to satisfy $\overline{\bm W} {\bm 1} = {\bm 1}$ afterwards. As the problem size considered is moderate $(n\leq 100)$, the network reconstruction problem \eqref{eq:nsi_l1} is solved using \texttt{cvx}.

We present the normalized mean square error (NMSE) under the above scenario in Fig.~\ref{fig:mse_vs_ns}. The NMSE of $\overline{\bm D}'$ is defined as $\| \hat{\bm D} - \overline{\bm D}' \|_F^2 / \| \overline{\bm D}'\|_F^2$ (and similarly for $\overline{\bm B}'$). The NMSE against $n_s$ is shown for two cases: (i) solving \eqref{eq:nsi_full} when ${\cal S} = \Omega_{\overline{\bm D}}$; (ii) solving \eqref{eq:nsi_l1} when ${\cal S} = [n-n_s] \times [n-n_s]$. We also include the NMSE curve when $\overline{\bm B}$ corresponds to a random ER bipartite graph with edge connectivity $p=0.1$.
The figure shows, first that only $n_s \approx 20$ stubborn agents are needed when the full support information is given. By contrast, we need $n_s \approx 40$ to attain a similar NMSE when there is no support information.
Comparing the NMSE to $d$-regular/random graph construction for $\overline{\bm B}$ shows that the recovery performance is significantly better when using the $d$-regular graph construction; e.g., if $d=5$, the NMSE of $\overline{\bm D}'$ is less than $10^{-3}$ with $n_s \geq 33$. This implies that by inserting almost half the number of ordinary agents into the network, the social network structure can be revealed with high accuracy. This result is consistent with the Theorem~\ref{thm:cs}, which predicts that when $\beta \geq 0.604$, i.e., $n_s \approx 36$, perfect recovery can be achieved. The discrepancy between the simulation results and Theorem~\ref{thm:cs} is possibly due to the fact that we are solving \eqref{eq:nsi_l1} instead of \eqref{eq:nsi_first2}. Moreover, in an ER graph, $\overline{\bm d}_i'$ is only $p (n-n_s)$-sparse on average, but Theorem~\ref{thm:cs} requires that \emph{every} row of $\overline{\bm D}'$ to be $p(n-n_s)$-sparse.

\begin{figure}[t]
\centering
\includegraphics[width =0.48\linewidth]{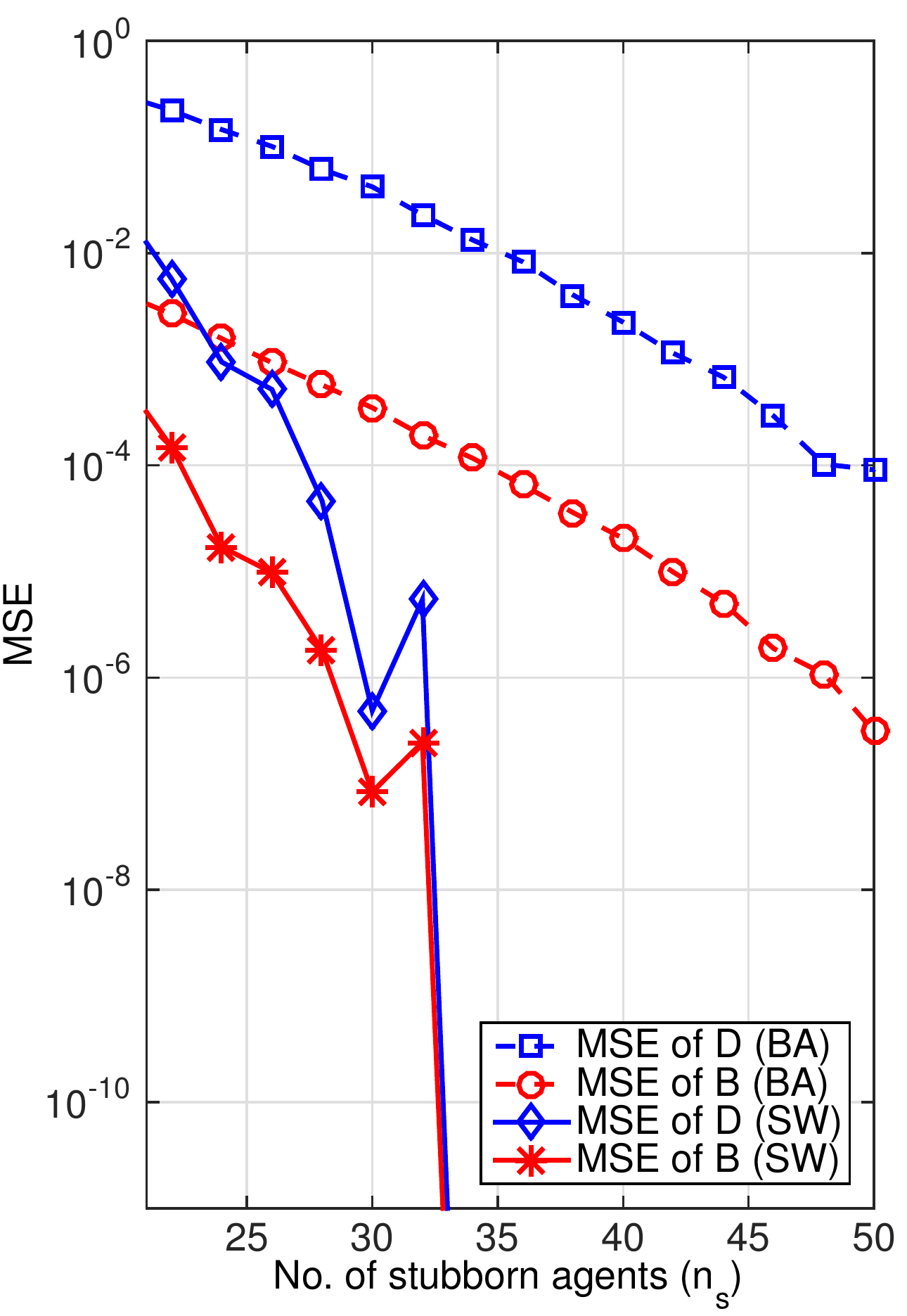} \includegraphics[width =0.48\linewidth]{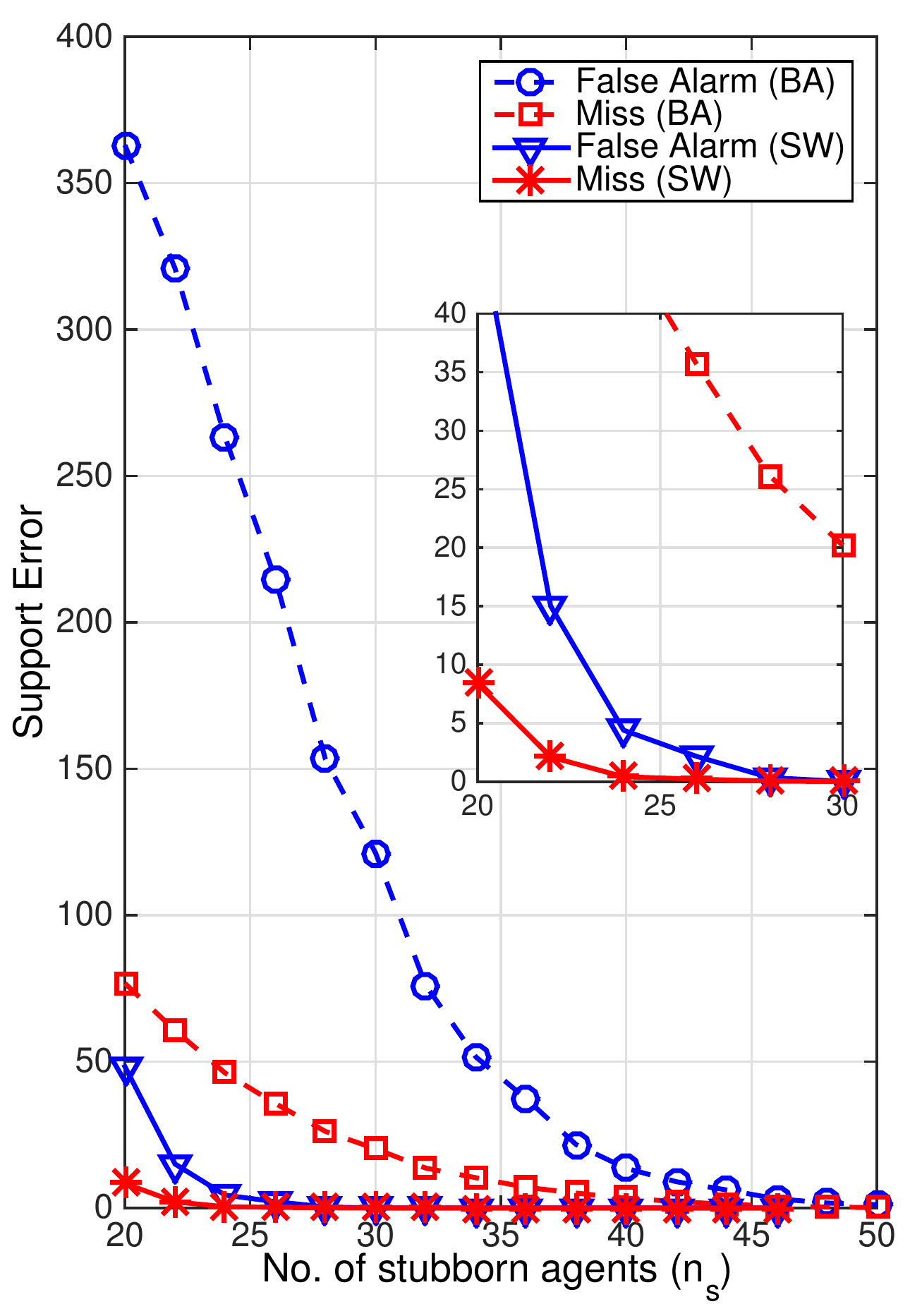}
\caption{Comparing performance against the number of stubborn agents $n_s$ with different network models. (Left) The NMSE. (Right) The average support recovery error.} \vspace{-.2cm}
\label{fig:mse_vs_ns_netwk}
\end{figure}

In the second example, we examine the scenarios when $G$ is constructed as the Barabasi-Albert (BA) graph {  with minimum degree $m=2$ for each incoming node}  \cite{Albert2002} or the Strogatz-Watts (SW) graph {  where each node is initially connected to $b=2$ left and right nodes and the rewiring probability is $p = 0.08$ }\cite{Watts1998}. The results are shown in Fig.~\ref{fig:mse_vs_ns_netwk}. It shows that the SW network can be recovered with high accuracy by using a much smaller number of stubborn agents than either the ER or BA networks. One possible explanation is that most of the vertices in SW have the same degree.

\begin{figure}[t]
\centering
\vspace{.4cm} \includegraphics[width =0.9\linewidth]{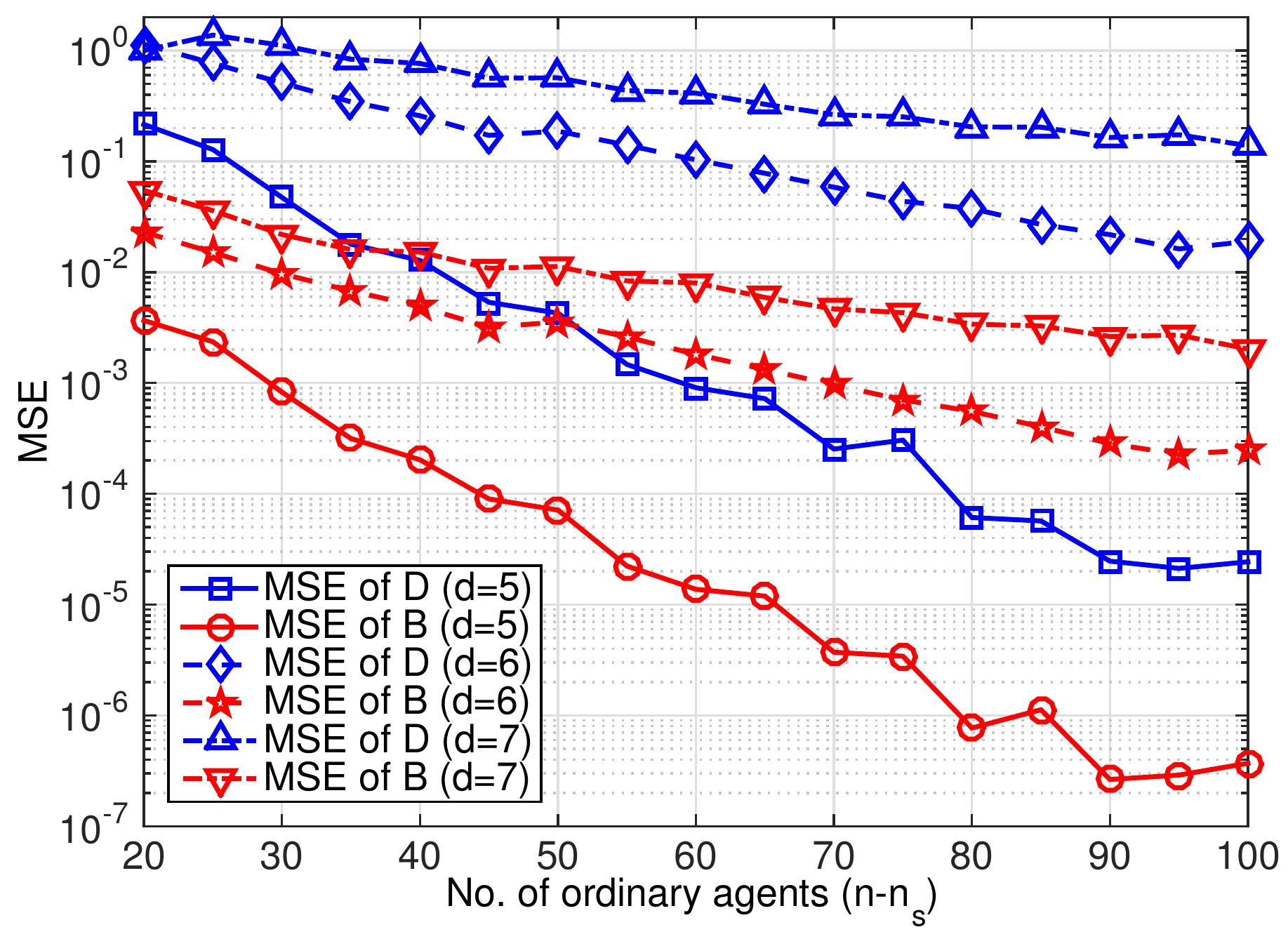} \vspace{-.0cm}
\caption{Comparing the NMSE against the number of ordinary agents $n-n_s$. Fix $p = 0.08$ and $\beta$ is given by Theorem~\ref{thm:cs} as $0.528 ~(d=5)$, $0.385 ~(d=6)$ and $0.319 ~(d=7)$.} \vspace{-.2cm}
\label{fig:mse_vs_n}
\end{figure}

The third numerical example examines the claim in Theorem~\ref{thm:cs}. Recall that the latter requires $n-n_s \rightarrow \infty$ for its validity. We consider the ER graph as in the first example and fix the connectivity of $G$ at $p=0.08$ where the smallest $\beta$ required by Theorem~\ref{thm:cs} are respectively $0.528 ~(d=5)$, $0.385 ~(d=6)$ and $0.319 ~(d=7)$. We set $n_s = \lceil \beta (n-n_s) \rceil$ and vary the number of ordinary agents $n-n_s$ to compare the  NMSE.
The NMSE comparison against $n-n_s$ can be found in Fig.~\ref{fig:mse_vs_n}. In all three cases tested $(d=5,6,7)$, there is a decreasing trend of the NMSE with $n-n_s$, suggesting that the failure probability decreases with $n-n_s \rightarrow \infty$. Moreover, although $d=7$ places the least requirement on $\beta$, it also has the highest probability of failure when $n$ is finite, since the upper bound to failure probability grows with ${\cal O}(d^5)$. The simulation results suggest that $d$ needs to be chosen judiciously when deciding on the required number/placement of stubborn agents.

\begin{figure}[t]
\centering
\vspace{.2cm}
\includegraphics[width =0.9\linewidth]{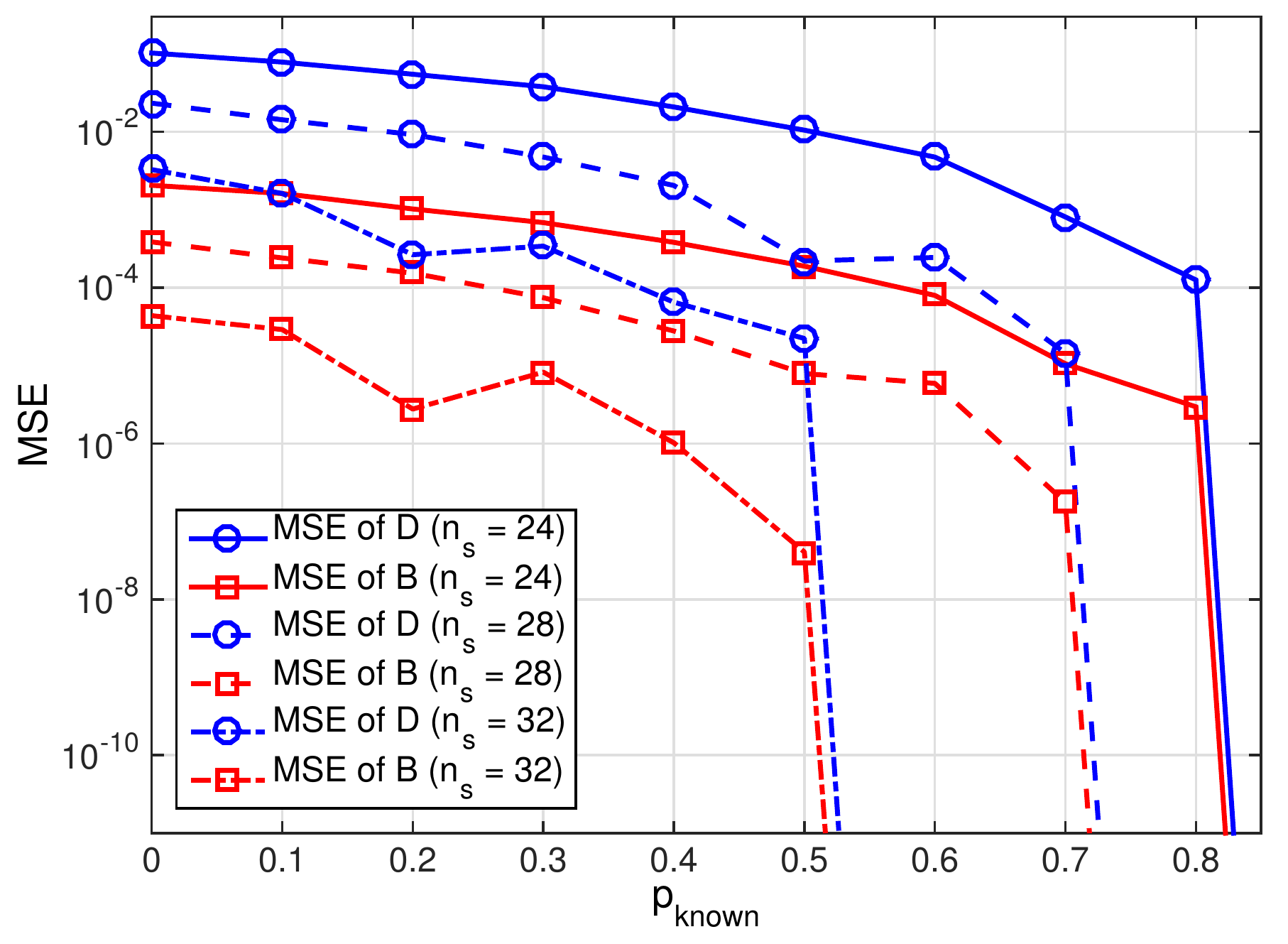} \vspace{-.3cm}
\caption{Comparing the NMSE against the percentage of known sparsity indices in $\Omega_{\overline{\bm W}}^c$, i.e., $|{\cal S}|$ decreases when $p_{known}$ increases .} \vspace{-.2cm}
\label{fig:sparsity}
\end{figure}

\begin{figure*}[ht]
\centering \vspace{-.2cm}
\includegraphics[width =0.49\linewidth]{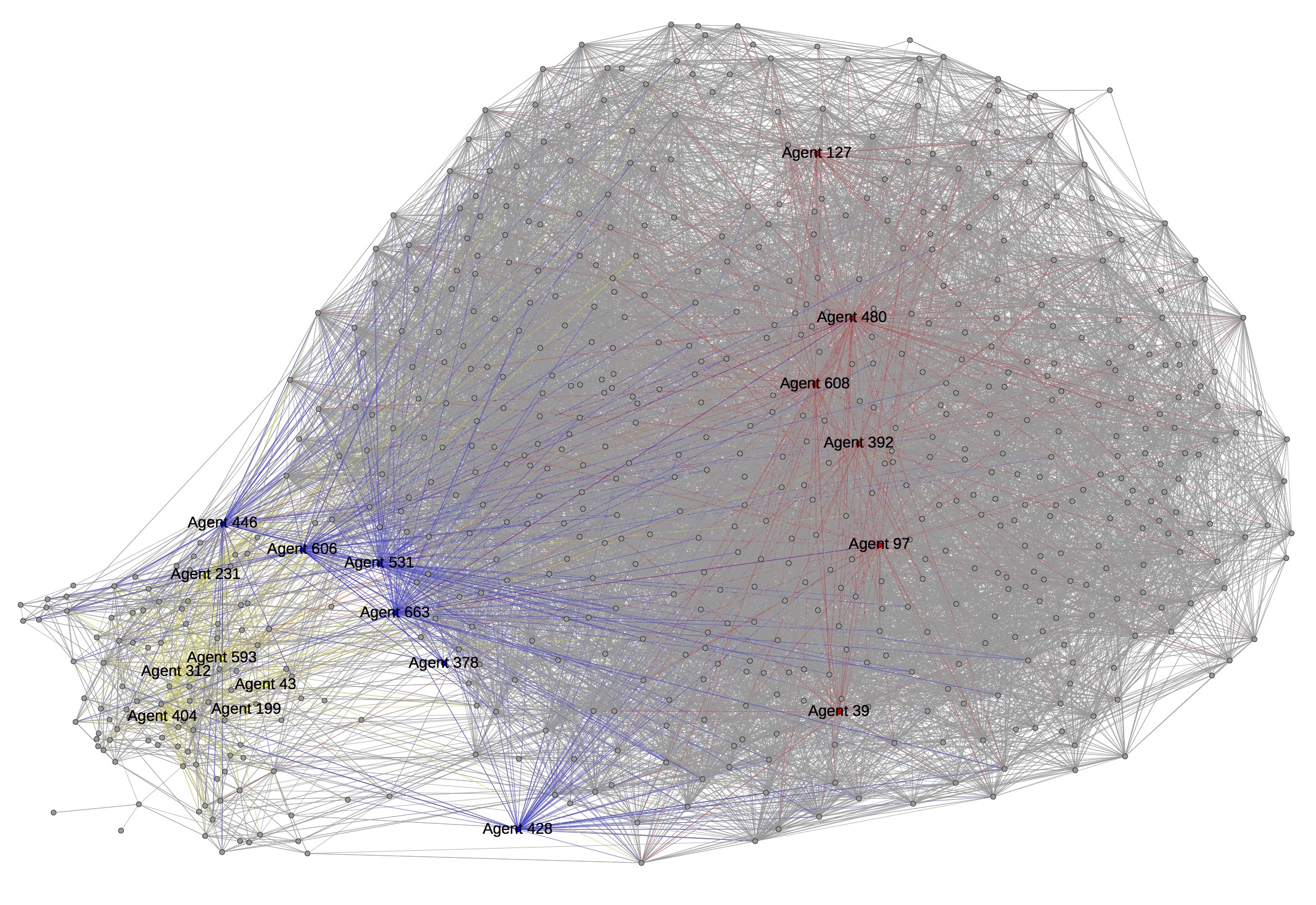}\includegraphics[width =0.49\linewidth]{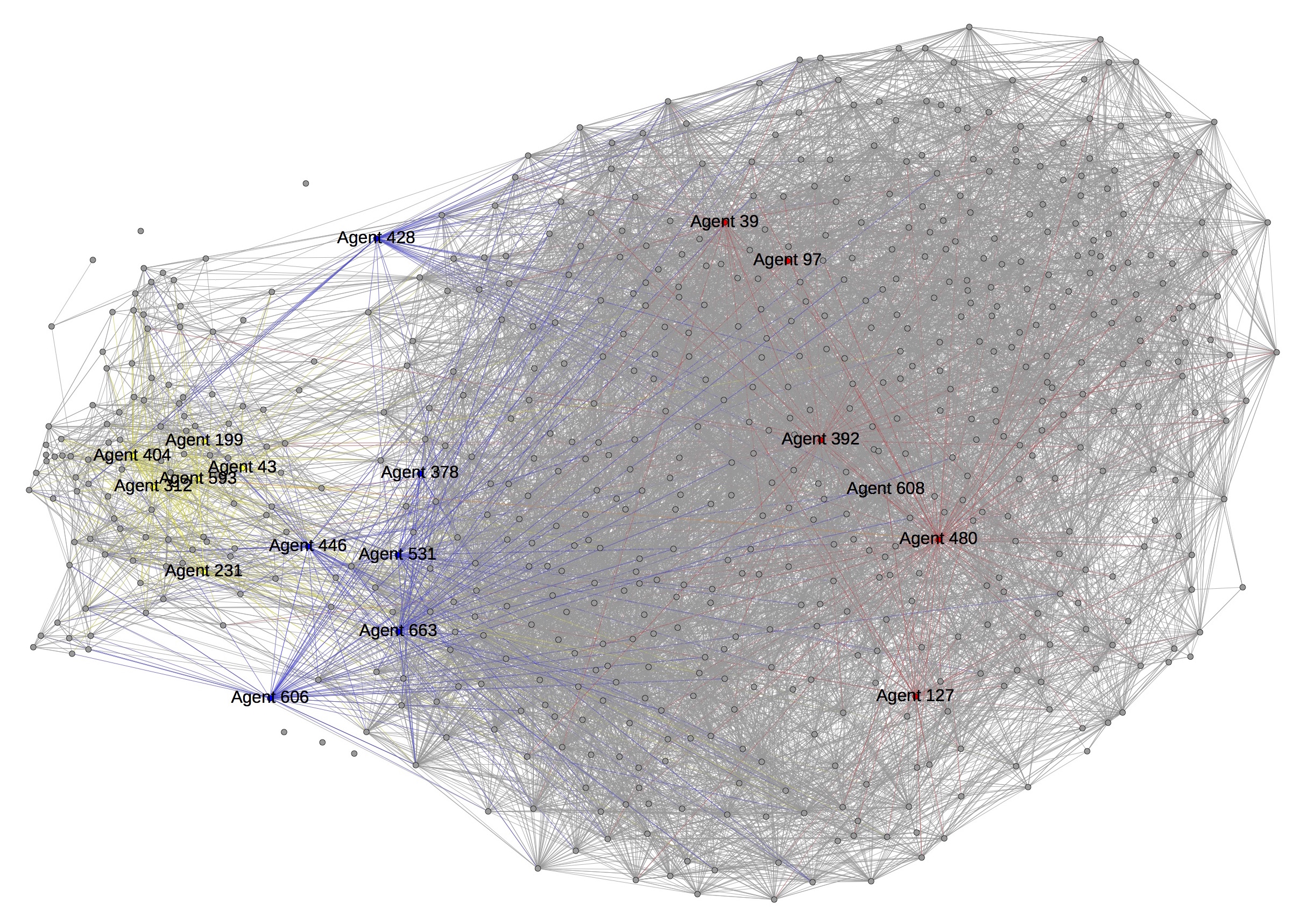} \vspace{-.3cm}
\caption{Comparing the social network of \texttt{ReedCollege} from the \texttt{facebook100} dataset: (Left) the original network; (Right) the estimated network.} \vspace{-.2cm}
\label{fig:facebook}
\end{figure*}

The next simulation example in Fig.~\ref{fig:sparsity} examines the case where a superset ${\cal S}$ of $\Omega_{\overline{\bm D}}$ is known. In particular, we consider the ER network model with $n-n_s=60$ and connectivity $p = 0.1$ and compare the NMSE against the percentage of exposed $\Omega_{\overline{\bm D}}^c$.
The figure shows that the network sensing performance improves as $p_{known}$ increases. When $40 \%$ of the support of $\overline{\bm D}$ is exposed, employing $n_s = 28$ stubborn agents yields a satisfactory NMSE of $10^{-3}$. \vspace{-.2cm}

\subsection{Real world networks}
This subsection examines the performance of the proposed method applied to real network data. Specifically, we consider the \texttt{facebook100} dataset \cite{Traud2012} and focus on the medium-sized network  example \texttt{ReedCollege}. The randomized opinion exchange model is based on the randomized broadcast gossip protocol in \cite{Aysal2009} with uniformly assigned trust weights.
Out of the available agents, we picked $n_s = 180$ agents with degrees closest to the median degree as the stubborn agents and removed the agents that are not adjacent to \emph{any} of the stubborn agents. {  The selection of the stubborn agents is motivated by Theorem~1 as we require a moderate average degree for the resultant stubborn-to-nonstubborn agent network with better recovery guarantees.}  Our aim is to estimate the trust matrix $\overline{\bm D}$, which corresponds to the subgraph with $n -n_s= 666$ ordinary agents, $|E| = 13,269$ edges and mean degree $19.92$. Note that the bipartite graph from stubborn agents to ordinary agents has a mean degree of $25.07$.

The opinion dynamics data $\hat{\bm Y}$ is collected using the estimator in Section~\ref{sec:random}, where we set $|{\cal T}_k| = 5 \times 10^5$ and the sampling instances are uniformly taken from the interval $[10^5, 5 \times 10^7]$.
We apply the FISTA algorithm developed in Section~\ref{sec:fast} to approximately solve the network reconstruction problem \eqref{eq:nsi_l1}, with $\lambda = n \times 10^{-12}$ and $\gamma = 10^{-3} /\gamma $. The NMSE of the reconstructed $\overline{\bm D}'$ is $0.1035$ after $4 \times 10^4$ iterations. The program has terminated in about $30$ minutes on an Intel\texttrademark~Xeon\texttrademark~server running MATLAB\texttrademark~2014b.
 It is computationally infeasible to deploy generic solvers such as \texttt{cvx} as the number of variables involved is $563,436$.

We compared the estimated social network in both macroscopic and microscopic levels. Fig.~\ref{fig:facebook} shows the true/estimated network plotted in \texttt{gephi} \cite{ICWSM09154} using the `Force Atlas 2' layout (with random initialization), where the edge weights are taken into account\footnote{Readers are advised to read the figures on a color version of this paper.}. While it is impossible to compare every edges in the network, the figure gives a macroscopic view of the efficacy of the network reconstruction method. In particular, using $n_s=180$ stubborn agents, the estimated network follows a similar topology as the original one. For instance, there are clearly two clusters in both the estimated and original network. Moreover, the relative roles for individual agents are matched in both networks. For example, agents $\{39,...,608\}$ are found in the larger cluster, agents $\{378,...,663\}$ are found at the boundary between the clusters and agents $\{43,...,404\}$ are found in the smaller cluster, in both networks.

\begin{figure}[t]
\centering
\includegraphics[width =0.49\linewidth]{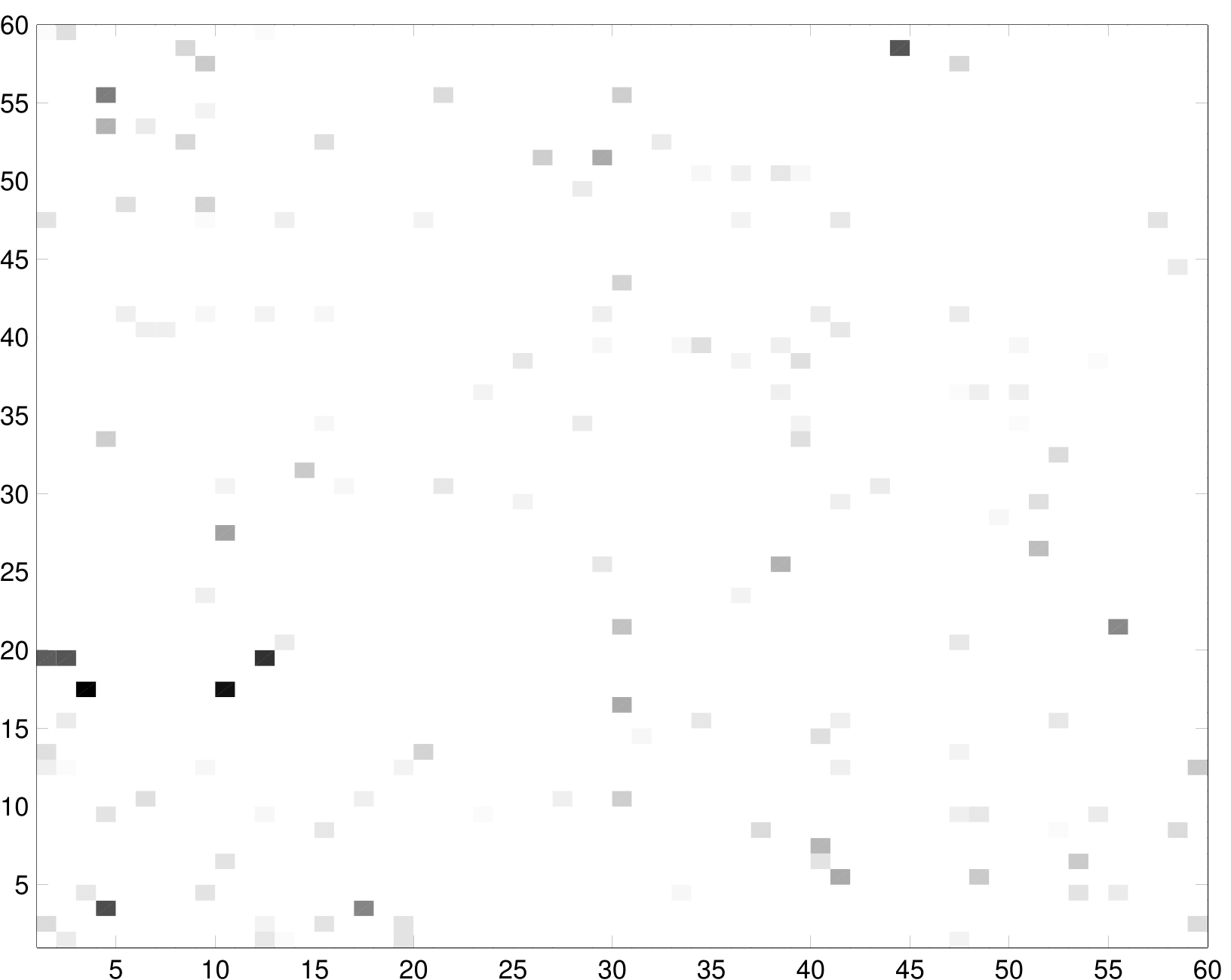}
\includegraphics[width =0.49\linewidth]{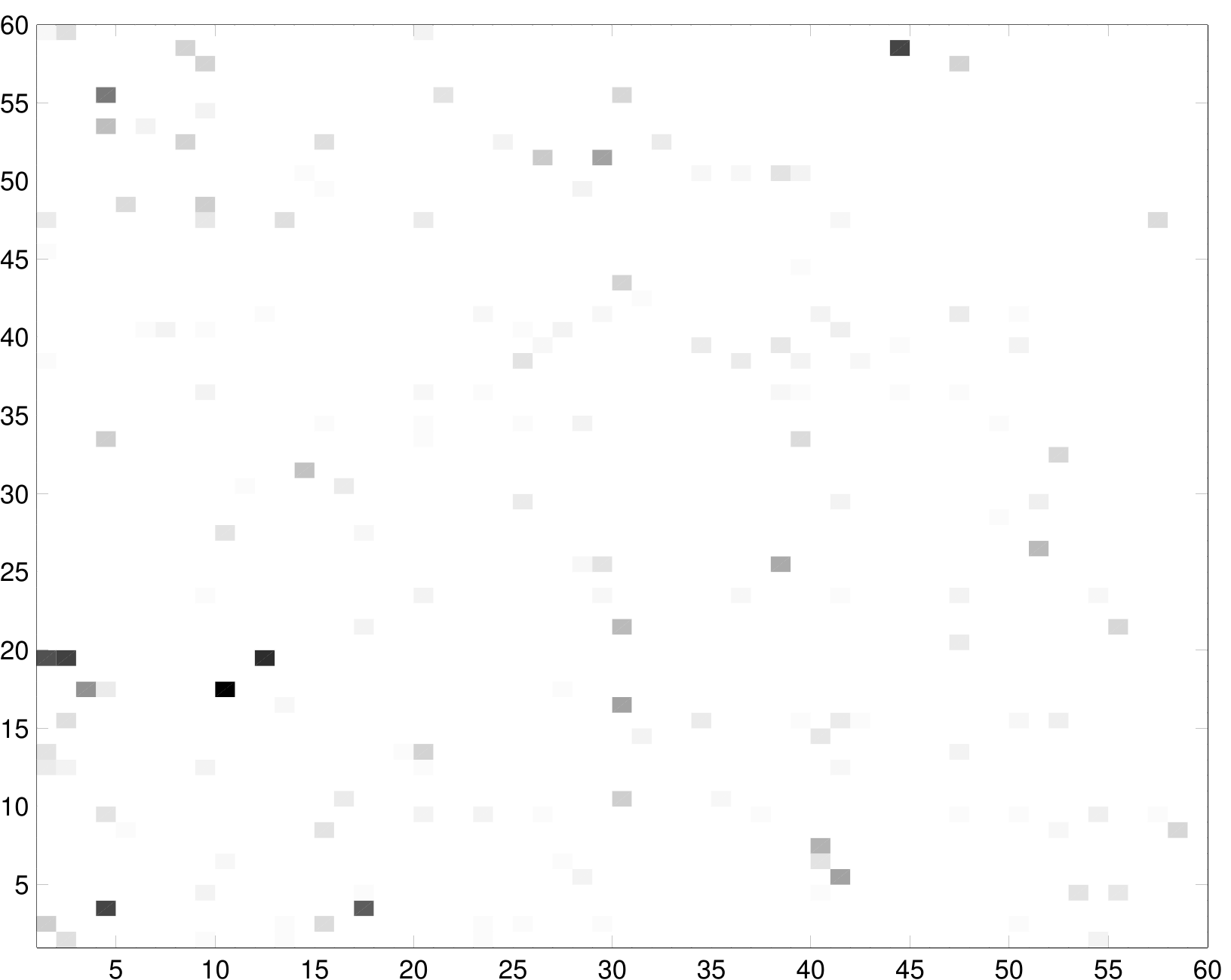}
\caption{Comparing the reconstructed network for the \texttt{ReedCollege} social network in \texttt{facebook100} dataset. (Left) Original network. (Right) Reconstructed network.} \vspace{-.2cm}
\label{fig:w_est}
\end{figure}

Finally, in Fig.~\ref{fig:w_est} we compare the estimated principal sub-matrix of $\overline{\bm D}'$ taken from the first 60 rows/columns, i.e., this corresponds to the social network between 60 agents. As seen, the original and estimated matrices appears to be similar to each other, both in terms of the support set and the weights on individual edges between the agents.

\section{Conclusions}

In this paper, we considered the social network sensing problem using data collected from steady states of opinion dynamics. The opinion dynamics is based on an extension of the linear DeGroot model with stubborn agents, where the latter plays a key role in exposing the network structure. Our main result is that the social network sensing problem can be cast as a sparse recovery problem and a sufficient condition for perfect recovery was proven. In addition, a consistent estimator was also derived to handle the case where the network dynamics is random and a low complexity algorithm is proposed. Our simulation results on synthetic and real networks indicate that the network structure can be reconstructed with high accuracy when a sufficient number of stubborn agents is present.

Ongoing research is focused on extending the active sensing method to nonlinear opinion dynamics models such as the Hegselmann and Krause model, working on real social network data collected from social media and combining this approach with the detection of stubborn agents.

\section*{Acknowledgement}
The authors are indebted to the anonymous reviewers for their invaluable comments to improve the current paper. 

\appendices

\section{Proof of Lemma~\ref{lem:amb}} \label{app:lemamb}
It is easy to check that:
\begin{equation}
\tilde{\bm B}{\bf 1} + \tilde{\bm D} {\bf 1} = \bm{\Lambda} ({\bm B} {\bf 1} + {\rm off}({\bm D}) {\bf 1}) + {\rm diag}(\tilde{\bm D}) = {\bf 1},
\end{equation}
where the last equality is due to ${\bm B} {\bm 1} + {\bm D} {\bm 1} = {\bm 1}$. Furthermore,
\begin{equation}
\begin{array}{l}
({\bm I} - \tilde{\bm D})^{-1}\tilde{\bm B} = ({\bm I} - \bm{\Lambda} {\rm off}({\bm D}) - {\rm Diag}({\rm diag}(\tilde{\bm D})) )^{-1}\bm{\Lambda}{\bm B}.
\end{array}
\end{equation}
As ${\rm Diag}({\rm diag}(\tilde{\bm D})) = {\bm I} - \bm{\Lambda} + \bm{\Lambda} {\rm Diag}({\rm diag}({\bm D}))$, we have:
\begin{equation}
\begin{array}{l}
({\bm I} - \tilde{\bm D})^{-1}\tilde{\bm B} = ({\bm I} - {\bm D})^{-1}{\bm B}.
\end{array}
\end{equation}

\section{Proof of Theorem~\ref{thm:cs}} \label{app:thmcs}
With a slight abuse of notations, in this appendix we assume that there are $n$ ordinary agents and $n_s$ stubborn agents. 
In particular, the dimensions of the variables are $\overline{\bm D}' \in \mathbb{R}^{n \times n}$ and $\overline{\bm B}' \in \mathbb{R}^{n \times n_s}$. 
For simplicity, we set $n_i = n$, $\alpha_i = \alpha$ and $\beta_i = \beta$ for all $i$. 

The proof of Theorem~\ref{thm:cs} is divided into two parts. The first part shows a sufficient condition for recovering $(\overline{\bm B}',\overline{\bm D}')$ using \eqref{eq:nsi_first2}; and the second part shows that the sufficient condition holds with high probability as $n \rightarrow \infty$.

\begin{figure}[t]
\centering
 \includegraphics[width =0.75\linewidth]{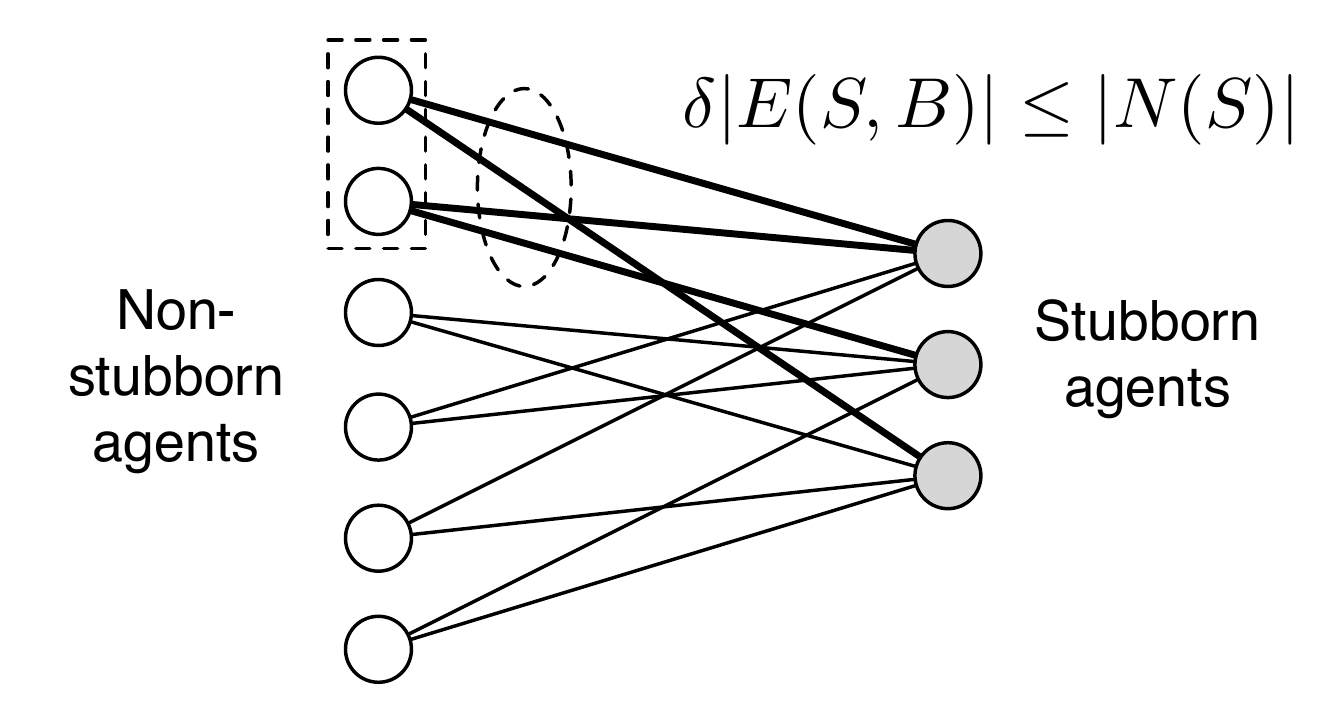} \vspace{-.2cm}
\caption{Illustrating the properties of the expander graph. In the above example bipartite graph, if $\alpha = 1/3$, $\delta$ is at most $3/4$ since $|E(S,B)| = 4$ and $|N(S)| = 3$ when $S$ is the first two vertices in the set of ordinary agents.} \vspace{-.2cm}
\label{fig:expander}
\end{figure}

Let $d(v)$ denote the degree of a vertex $v$. 
Our proof relies on the following definition of an unbalanced expander graph:
\begin{Def}
An $(\alpha,\delta)$-unbalanced expander graph is an $A,B$-bipartite graph (bigraph) with $|A| = n, |B| = m$ with left degree bounded in $[d_l, d_u]$, i.e., $d(v_i) \in [d_l, d_u]$ for all $v_i \in A$, such that for any $S \subseteq A$ with $|S| \leq \alpha n$, we have $\delta |E(S,B)| \leq |N(S)|$, where $E(S,B)$ is the set of edges connected from  $S$ to $B$ and $N(S) = \{ v_j \in B : \exists~v_i \in S~s.t.~v_j v_i \in E \}$ is the neighbor set of $S$ in $B$.
\end{Def}
We imagine that $A$ ($B$) is the set of ordinary (stubborn) agents and $E(A,B)$ represents the connection between stubborn and ordinary agents; see the  illustration in Fig.~\ref{fig:expander}.
We denote the collection of $(\alpha,\delta)$-unbalanced expander graphs by ${\cal G}(\alpha,\delta)$.
Previous works \cite{Berinde2008,Wang2011,Khajehnejad2011,Gilbert2010} have shown that the expander graph structure allows for the construction of measurement matrices with good sparse recovery performance.

We now proceed by showing the sufficient condition.
Denote the support of ${\bm b}_i - \overline{\bm b}_i'$ as $\Omega_{\overline{\bm B}}^i$, where $|\Omega_{\overline{\bm B}}^i| = d$. Since $\Omega_{\overline{\bm B}}^i$ is known a-priori, ${\bm b}_i - \overline{\bm b}_i'$ is a sparse vector supported on $\Omega_{\overline{\bm B}}^i$. We can thus treat the rows where ${\bm b}_i$ is supported on as `erasure bits', which can be ignored. In particular, the following rows-deleted linear system can be deduced from the last line in \eqref{eq:linear_rel}:
\begin{equation} \label{eq:row_del}
\overline{\bm B}_{(\Omega_{\overline{\bm B}}^i)^c}^{'T} ({\bm I} - \overline{\bm D}')^{-T}  (\overline{\bm d}_i' - {\bm d}_i ) = {\bm 0},
\end{equation}
where $\overline{\bm B}_{(\Omega_{\overline{\bm B}}^i)^c}^{'T}$ is a $d$-rows-deleted matrix obtained from $\overline{\bm B}^{'T}$.

We prove the sufficient condition by deriving a Restricted Isometry Property-1 (RIP-1) condition for ${\bm A} = {\bm B}_{(\Omega_{\overline{\bm B}^i})^c}^T$ and its perturbation ${\bm A} ({\bm I} - \overline{\bm D}')^{-T}$. We define $a_{min} = \min_{ij \in {\rm supp}({\bm A})} A_{ij}$ and $a_{max} = \max_{ij \in {\rm supp}({\bm A})} A_{ij}$ and prove the following proposition:
\begin{Prop}\label{lem:rip1}
Let $n > m$ and ${\bm A} \in \mathbb{R}^{m \times n}$ be a non-negative matrix that has the same support as the adjacency matrix of an $(\alpha,\delta)$-unbalanced bipartite expander graph with bounded left degrees $[d_l, d_u]$.
Then ${\bm A}$ satisfies the RIP-1 property:
\begin{equation} \label{eq:rip_1}
\big( a_{min} \delta d_l -  a_{max} (d_u-\delta d_l) \big)  \| {\bm x} \|_1 \leq \| {\bm A} {\bm x} \|_1 \leq d_u a_{max} \| {\bm x} \|_1,
\end{equation}
for all $k$-sparse ${\bm x}$ such that $k \leq \alpha n$. Furthermore, we have
\begin{equation} \label{eq:rip_inverse}
\upsilon^\star \cdot \| {\bm x} \|_1 \leq \| {\bm A} ({\bm I} - \overline{\bm D}')^{-T} {\bm x} \|_1,
\end{equation}
where $\upsilon^\star = a_{min} \delta d_l -  a_{max} (d_u-\delta d_l) - (1 - d_l a_{min} )$.
\end{Prop}
\noindent \textbf{Proof}. The following proof is a generalization of \cite[Appendix D]{Khajehnejad2011}. First of all, the upper bound in \eqref{eq:rip_1} follows from $\| {\bm A} {\bm x} \|_1 \leq \| {\bm A} \|_{1,1} \|{\bm x}\|_1$, where $\| {\bm A} \|_{1,1}$ is the matrix norm induced by $\| \cdot \|_1$ on ${\bm A}$ \cite{Horn1985}, i.e.,
\begin{equation}
\| {\bm A} \|_{1,1} = \max_{ 1 \leq j \leq n } \sum_{i=1}^{m} |A_{ij}|.
\end{equation}
Obviously we have $\| {\bm A} \|_{1,1} \leq d_u a_{max}$.

To prove the lower bound in \eqref{eq:rip_1}, using the expander property, we observe that
\begin{equation}
\delta d_l |S| \leq \delta |E(S,B)| \leq |N(S)|,
\end{equation}
for all $S \subseteq {\rm supp}({\bm x}) = \{ i : x_i \neq 0\}$ and $|S| \leq \alpha n$. As a consequence of Hall's theorem \cite{west_introduction_2000}, the bigraph induced by ${\bm A}$ contains $\delta d_l$  disjoint matchings for ${\rm supp}({\bm x})$. We can thus decompose ${\bm A}$ as:
\begin{equation}
{\bm A} = {\bm A}_M + {\bm A}_C,
\end{equation}
where the decomposition is based on dividing the support such that ${\rm supp}({\bm A}_M) \cap {\rm supp}({\bm A}_C) = \emptyset$. In particular, ${\bm A}_M$ is supported on the $\delta d_l$ matchings for ${\rm supp}({\bm x})$; i.e., by the matching property, each row of ${\bm A}_M$ has at most one non-zero, and each column of ${\bm A}_M$ has $\delta d_l$ non-zeros, and the remainder ${\bm A}_C$ has at most $d_u -\delta d_l$ non-zeros per column. Applying the triangular inequality gives:
\begin{equation}
\| {\bm A} {\bm x} \|_1 \geq \| {\bm A}_M {\bm x} \|_1 - \| {\bm A}_C {\bm x} \|_1,
\end{equation}
since $\|{\bm A}_M {\bm x}\|_1 \geq a_{min} \delta d_l \|{\bm x}\|_1$ and $\|{\bm A}_C {\bm x} \|_1 \leq  a_{max} (d_u-\delta d_l) \|{\bm x}\|_1$, this implies:
\begin{equation} \label{eq:2ndlast}
\| {\bm A} {\bm x} \|_1 \geq \big( a_{min} \delta d_l -  a_{max} (d_u-\delta d_l) \big) \| {\bm x} \|_1.
\end{equation}
For the second part in the lemma, i.e.,  \eqref{eq:rip_inverse}, note that:
\begin{equation} \label{eq:lasteq}
\| {\bm A} ({\bm I} - \overline{\bm D}')^{-T} {\bm x} \|_1 \geq \| {\bm A} {\bm x} \|_1 - \| {\bm A} \overline{\bm D'}^{T} ({\bm I} - \overline{\bm D}')^{-T} {\bm x} \|_1,
\end{equation}
since ${\bm A} ({\bm I} - \overline{\bm D}')^{-T} {\bm x} = {\bm A} {\bm x} + {\bm A} \overline{\bm D}'^T({\bm I} - \overline{\bm D}')^{-T} {\bm x}$.
The latter quantity can be upper bounded by
\begin{equation}\label{eq:easy_bd}
\hspace{-.1cm} \begin{array}{l}
\| {\bm A} \overline{\bm D'}^{T} ({\bm I} - \overline{\bm D}')^{-T} {\bm x} \|_1 \\
~\leq \|{\bm A}\|_{1,1} \| \overline{\bm D'}^{T} \|_{1,1} \| ({\bm I} - \overline{\bm D}')^{-T} \|_{1,1} \|{\bm x}\|_1 \\
\displaystyle ~\leq d_u a_{max} \frac{ \| \overline{\bm D'}^{T} \|_{1,1} }{ 1 - \| \overline{\bm D'}^{T} \|_{1,1} } \|{\bm x} \|_1 \leq (1 - d_l a_{min} ) \|{\bm x} \|_1,
\end{array}\hspace{-.2cm}
\end{equation}
where in the second to last inequality, we 	used the property $\| ( {\bm I} - {\bm C} )^{-1} \| \leq 1 / (1 - \| {\bm C} \| )$ for any $\| {\bm C}\| < 1$ \cite{Horn1985}; and in the last inequality, we used the fact that $1 - d_u a_{max} \leq \| \overline{\bm D'}^{T} \|_{1,1} \leq 1 - d_l a_{min}$ (note that each row in $\overline{\bm D'}$ sums to at most $1 - d_l a_{min}$ and at least $1 - d_u a_{max}$). Combining \eqref{eq:2ndlast}, \eqref{eq:lasteq} and \eqref{eq:easy_bd} yields the desired inequality. \hfill {\bf Q.E.D.}

A sufficient condition for $\ell_0$ recovery can be obtained by proving the following simple corollary:
\begin{Corollary}\label{lem:suff}
Let the conditions from Proposition~\ref{lem:rip1} on ${\bm A}$ holds. Suppose that both ${\bm x}_1,{\bm x}_2$ are $(k/2)$-sparse such that $k \leq \alpha n$ and:
\begin{equation}
{\bm A} ({\bm I} - \overline{\bm D}')^{-T} {\bm x}_1 = {\bm A} ({\bm I} - \overline{\bm D}')^{-T} {\bm x}_2,
\end{equation}
then ${\bm x}_1 = {\bm x}_2$ if
\begin{equation}
\upsilon^\star = a_{min} \delta d_l -  a_{max} (d_u-\delta d_l) - (1 - d_l a_{min} ) > 0.
\end{equation}
\end{Corollary}
\noindent \textbf{Proof}. Observe that ${\bm x}_1 - {\bm x}_2$ is at most $k$-sparse, using Proposition~\ref{lem:rip1}, we have
\begin{equation}
\upsilon^\star \|{\bm x}_1 - {\bm x}_2 \|_1 \leq \| {\bm A} ({\bm I} - \overline{\bm D}')^{-T} ({\bm x}_1 - {\bm x}_2) \|_1 = 0.
\end{equation}
This implies that ${\bm x}_1 = {\bm x}_2$. \hfill {\bf Q.E.D.}

As $\overline{\bm d}_i'$ is $k/2$-sparse, $b_{min} \leq a_{min}$ and $b_{max} \geq a_{max}$, Eq.~\eqref{eq:thmval} and Corollary~\ref{lem:suff} guarantee that $\overline{\bm d}_i'$ is the \emph{unique} solution out of all $k/2$-sparse vectors that ${\bm d}_i$ satisfies  \eqref{eq:row_del}. This means that any ${\bm d}_i$ that satisfies \eqref{eq:row_del} must be either $\overline{\bm d}_i'$ or have $\| {\bm d}_i \|_0 > (k/2)$. Since the optimization problem \eqref{eq:nsi_first2} finds the sparsest solution satisfying \eqref{eq:row_del}, we must have ${\bm d}_i^\star = \overline{\bm d}_i'$ for all $i$. Furthermore, this implies ${\bm b}_i^\star = \overline{\bm b}_i'$ in \eqref{eq:linear_rel} and we have $({\bm B}^\star, {\bm D}^\star) = (\overline{\bm B}, \overline{\bm D}')$.

The second part of our proof shows that for all $i$, the support set of the $d$-rows-deleted matrix $\overline{\bm B}_{(\Omega_{\overline{\bm B}}^i)^c}^{'T}$ corresponds to an $(\alpha,\delta)$-expander graph with high probability. Our plan is to first prove that the corresponding bipartite graph has a bounded degree $r \in [d-1,d]$ with high probability (w.h.p.), and then show that a randomly constructed bipartite with bounded degree $r \in [d-1,d]$ is also an expander graph w.h.p..

Now, let us observe the following proposition:
\begin{Prop} \label{prop:row}
Let $G$ be a random $A,B$-bigraph with $|A| = n$, $|B| =n_s =  \beta n$, constructed by randomly connecting $d$ vertices from $A$ to each vertex of $B$. All of the subgraphs $G_1,...,G_n$ have left degree $r \in [d-1,d]$ with high probability (as $n \rightarrow \infty$) if each of these subgraphs is formed by randomly deleting $d$ vertices from $B$ in $G$.
\end{Prop}

\noindent \textbf{{Proof.}} We lower bound the desired probability as follows:
\begin{equation} \label{eq:firstchain}
\begin{array}{l}
{\rm Pr} \Big( G_1,...,G_n = \text{bipartite with (left) deg.~$r \in [d-1,d]$}\Big) \\
= 1 - {\rm Pr} \Big( \cup_{i=1}^n (G_i = \text{bipartite with min.~deg.~$r < d - 1$} )\Big) \\
\geq 1 - n \cdot {\rm Pr} \Big( \cup_{k=1}^n (d(v_k) < d-1,~v_k \in A_i, ~A_i \subseteq V(G_i) ) \Big) \\
\geq 1 - n^2 \cdot {\rm Pr} \Big( d(v_k) < d-1,~v_k \in A_i, ~A_i \subseteq V(G_i) \Big)
\end{array}
\end{equation}
Note that the event described in the last term is equivalent to deleting at least $2$ neighbors of $v_k \in A_i$ from $B$. As the neighbors of $A$ are also randomly selected, the latter probability can be upper bounded by:
\begin{equation}
\begin{array}{l}
{\rm Pr} \Big( d(v_k) < d-1,~v_k \in A_i, ~A_i \subseteq V(G_i) \Big)\\
\displaystyle = {\rm Pr} \big( d(v_k) =0 \cup \cdots \cup d(v_k) = d-2 \big) \\
\displaystyle \leq (d-1) \cdot \left( \frac{ d^2 }{ (\beta n)^2 }\right)^2 = (d-1) \cdot \left( \frac{d}{\beta n} \right)^4,
\end{array}
\end{equation}
Plugging this back into \eqref{eq:firstchain} yields the desired result. \hfill {\bf Q.E.D.}

The proof of Theorem~\ref{thm:cs} is completed by the proposition:
\begin{Prop} \label{prop:exp}
Let $G$ be a random $A,B$-bigraph with $|A| = n$, $|B| = \beta' n = n_s - d$, constructed by randomly connecting $r \in [d-1,d]$ vertices from $A$ to each vertex of $B$. Then $G$ is an $(\alpha, 1-1/(d-1))$-expander graph with high probability if $d \geq 4$, $\alpha < \beta'$ and $d - 1> (H(\alpha) + \beta' H(\alpha/\beta') ) / \alpha \log( \beta' / \alpha)$.
\end{Prop}

\noindent \textbf{{Proof.}} The following proof is similar in flavor to the proof of  \cite[Proposition~1]{Khajehnejad2011}, with the additional complexity that the left degree is variable. For simplicity, we denote ${\bm A}$ as the adjacency matrix of $G$ and let $E_{i_1,...,i_r}$ be the event such that ${\bm A}_{:,i_1,...,i_r}$ contains at least $m-r+1$ zero rows, where ${\bm A}_{:,i_1,...,i_r}$ is the submatrix formed by choosing the $\{i_1,...,i_r\}$ columns. Note that if $r \leq \alpha n$ and $E_{i_1,...,i_r}$ occurs, $G \notin {\cal G} (\alpha, 1 - 1/(d-1))$ since $(1 - 1/(d-1)) |E( \{ i_1,...,i_r \} )| \geq r > r-1 = |N(\{ \{ i_1,...,i_r \})|$. The failure probability can thus be upper bounded as:
\begin{equation}
\begin{array}{l}
\displaystyle {\rm Pr} \Big( G \notin {\cal G}(\alpha,1-1/(d-1)) \Big) \\
\displaystyle \leq {\rm Pr} \Big( \bigcup_{d-1\leq r \leq \alpha n, 1\leq i_1 < i_2 < \cdots < i_r} E_{i_1,...,i_r} \Big) \\
\displaystyle \leq \sum_{r=d-1}^{\alpha n} {n \choose r} {\rm Pr} ( E_{1,...,r} ).
\end{array}
\end{equation}

Suppose that there are $r-s$ columns with $d-1$ non-zero entries and $s$ columns with $d$ non-zero entries; hence we have ${\beta n \choose d-1}^{r-s} {\beta n \choose d}^s$
possible sub-matrices to choose from. Now, a necessary condition for $E_{1,...,r}$ is such that all the non-zero entries are contained in a sub-sub-matrix of size $r \times r$. There are at most ${r \choose d-1}^{r-s} {r \choose d}^s$
possible configurations and ${\beta n \choose r}$ such sub-sub-matrices. For this case, we obtain the upper bound:
\begin{equation}
\begin{split}
{\rm Pr}(E_{1,...,r},~{\rm fix}~s) & \leq \frac{ {\beta' n \choose r} {r \choose d-1}^{r-s} {r \choose d}^s }{ {\beta' n \choose d-1}^{r-s} {\beta' n \choose d}^s } \\
& \leq {\beta' n \choose r} \cdot \left( \frac{  r }{ \beta' n }\right)^{(r-s)(d-1)} \cdot \left( \frac{  r }{ \beta' n }\right)^{sd} \\
& = {\beta' n \choose r} \cdot \left( \frac{  r }{ \beta' n }\right)^{(r-s)(d-1) + sd}
\end{split}
\end{equation}
where we used the fact that ${r \choose d} / {m \choose d} \leq (r/m)^d$ if $r < m$. Taking the union bound for all configurations {  $s \in [0,r]$} gives:
\begin{equation}
\begin{array}{l}
{\rm Pr}(E_{1,...,r}) \leq \sum_{s=0}^r {\rm Pr}(E_{1,...,r},~{\rm fix}~s) \vspace{.1cm} \\
\displaystyle \leq  {\beta' n \choose r} \cdot \left( \left( \frac{  r }{ \beta' n }\right)^{r(d-1)}  \hspace{-.3cm} + \left( \frac{  r }{ \beta' n }\right)^{r(d-1)+1}  \hspace{-.6cm} + \cdots +  \left( \frac{  r }{ \beta' n }\right)^{rd}  \right) \\
\displaystyle = {\beta' n \choose r} \cdot \frac{1}{1 - r/(\beta' n)} \left(  \left(\frac{  r }{ \beta' n }\right)^{r(d-1)} - \left( \frac{  r }{ \beta' n }\right)^{rd+1} \right) \\
\displaystyle < \frac{1}{1 - \alpha/\beta'} {\beta' n \choose r} \cdot \left(\frac{  r }{ \beta' n }\right)^{r(d-1)}
\end{array}
\end{equation}
The second equality is due to the geometric series and the last inequality is due to $r \leq \alpha n$. We thus have:
\begin{equation}
\begin{array}{l}
{\rm Pr} \Big( G \notin {\cal G}(\alpha,1-1/(d-1))  \Big) \\
~~~~~~~\displaystyle \leq \frac{1}{1 - \frac{\alpha}{\beta'}} \sum_{r=d-1}^{\alpha n} {n \choose r} {\beta' n \choose r} \left(\frac{  r }{ \beta' n }\right)^{r(d-1)}
\end{array}
\end{equation}
The remainder of the proof follows from that of \cite{Khajehnejad2011}; i.e., Lemma~A.1 and A.2, through replacing $d$ by $d-1$.
In particular, we can show that
\begin{equation}
{\rm Pr} \Big( G \notin {\cal G}(\alpha,1-1/(d-1))  \Big)  \leq {\cal O}(n^{1-(d-1)(d-3)})
\end{equation} if $d - 1> (H(\alpha) + \beta' H(\alpha/\beta') ) / \alpha \log( \beta' / \alpha)$. This completes the proof. \hfill {\bf Q.E.D.}

Combining Proposition~\ref{prop:row} \& \ref{prop:exp} indicates that the $d$-rows-deleted sensing matrix $\overline{\bm B}_{(\Omega_{\overline{\bm B}}^i)^c}^{'T}$ corresponds to an $(\alpha, 1-1/(d-1))$-expander graph with high probability. Therefore, the conclusion in Corollary~\ref{lem:suff} follows by setting $\delta = 1 - 1/(d-1)$. Moreover, by applying the union bound, the probability of failure is upper bounded as:
 \begin{equation}
{\rm Pr}( Fail ) \leq \left( \frac{d}{\beta} \right)^4 \frac{d-1}{n^2} + {\cal O}( n^{2 - (d-1)(d-3)} ),
\end{equation}
which vanishes as $n \rightarrow \infty$.

\section{Proof of Theorem~\ref{thm:asymp}} \label{app:cons}
To simplify the notations, in this section, we  drop the dependence on the discussion index $k$ for the opinion vectors ${\bm x}(t;k)$ and the trust matrices ${\bm W}(t;k)$.
We first prove that the estimator is unbiased. Consider the following chain:
\begin{equation}
\begin{array}{l}
\displaystyle \mathbb{E} \{ \hat{\bm x}({\cal T}_k ) | {\bm x}(0) \} = \frac{1}{|{\cal T}_k|} \sum_{t_i \in {\cal T}_k} \mathbb{E} \{ \hat{\bm x}(t_i) | {\bm x}(0) \} \\
\displaystyle ~~=  \frac{1}{|{\cal T}_k|} \sum_{t_i \in {\cal T}_k} \overline{\bm W}^{t_i} {\bm x}(0) = \overline{\bm W}^\infty {\bm x}(0),
\end{array}
\end{equation}
where we  used the fact that $T_o \rightarrow \infty$ and $t_i \geq T_o$ for all $t_i$ in the last equality.

Next, we prove that the estimator is asymptotically consistent, i.e., \eqref{eq:est_cons}. Without loss of generality, we let $t_1 < t_2 < \ldots < t_{|{\cal T}_k|}$ as the sampling instances. The following shorthand notation will be useful:
\begin{equation} \label{eq:phi_def}
\bm{\Phi}(s,t) \triangleq \bm{W}(t) \bm{W}(t-1) \ldots \bm{W}(s+1) \bm{W}(s),
\end{equation}
where $t \geq s$ and $\bm{\Phi}(s,t)$ is a random matrix.
Our proof involves the following lemma:
\begin{Lemma} \label{lemma:phi}
When $|t-s| \rightarrow \infty$, the random matrix $\bm{\Phi}(s,t)$ converges almost surely to the following:
\begin{equation}
\lim_{ |t-s| \rightarrow \infty} \bm{\Phi}(s,t) = \left(
\begin{array}{cc}
{\bm I} & {\bm 0} \\
{\bm B}(s,t) & {\bm 0}
\end{array}
\right),
\end{equation}
where ${\bm B}(s,t) = \sum_{q=s}^t ( {\bm D}(t) \ldots {\bm D}(q) ) {\bm B}(q)$. Moreover, ${\bm B}(s,t) $ is bounded almost surely.
\end{Lemma}

\emph{Proof:}  We first establish the almost sure convergence of ${\bm D}(t) {\bm D}(t-1) \ldots {\bm D}(s)$ to ${\bf 0}$. Define
\begin{equation}
\beta(s,t) \triangleq \| {\bm D}(t) {\bm D}(t-1) \ldots {\bm D}(s) \|_2,
\end{equation}
and observe the following chain
\begin{equation}
\begin{array}{l}
\mathbb{E} \{ \beta(s,t) | \beta(s,t-1),...,\beta(s,s) \} \vspace{.05cm} \\
 = \mathbb{E} \{ \| {\bm D}(t) {\bm D}(t-1) \ldots {\bm D}(s) \|_2 | \beta(s,t-1) \}  \vspace{.05cm} \\
  \leq \mathbb{E} \{ \| {\bm D}(t) \|_2 \| {\bm D}(t-1) \ldots {\bm D}(s) \|_2 | \beta(s,t-1) \}  \vspace{.05cm} \\
  = \mathbb{E} \{ \| {\bm D}(t) \|_2 \} \beta(s,t-1) \leq c \beta(s,t-1),
\end{array}
\end{equation}
where $c = \| \overline{\bm D} \|_2 < 1$. The almost sure convergence of $\beta(s,t)$ follows from \cite[Lemma~7]{polyak87}. Now, expanding the multiplication \eqref{eq:phi_def} yields:
 \begin{equation}
\bm{\Phi}(s,t) = \left(
\begin{array}{cc}
{\bm I} & {\bm 0} \\
{\bm B}(s,t) & {\bm D}(t) \ldots {\bm D}(s)
\end{array}
\right).
\end{equation}
The desired result is achieved by observing ${\bm D}(t) \ldots {\bm D}(s) \rightarrow {\bf 0}$ as $|t-s| \rightarrow \infty$. Lastly, the almost sure boundedness of ${\bm B}(s,t)$ can be obtained from the fact that $\bm{\Phi}(s,t)$ is stochastic. 
\hfill \textbf{Q.E.D.}

We consider the following chain:
\begin{equation}
\begin{array}{l}
\displaystyle \mathbb{E} \{ \| \hat{\bm x}({\cal T}_k) - \overline{\bm x}(\infty) \|_2^2 | {\bm x}(0) \} = \vspace{.1cm} \\
\displaystyle ~~= \mathbb{E} \Big\{ \Big\| \frac{1}{|{\cal T}_k|} \sum_{t_i \in {\cal T}_k} \big( \hat{\bm x}(t_i) - \overline{\bm x}(\infty) \big) \Big\|_2^2 | {\bm x}(0) \Big\}.
\end{array}
\end{equation}
Recall that $\hat{\bm x}(t_i) = {\bm x}(t_i) + {\bm n}(t_i)$ and the noise term ${\bm n}(t_i)$ is independent of ${\bm W}(t)$ for all $t$. The above expression reduces to:
\begin{equation}
\begin{array}{l}
 \mathbb{E} \Big\{ \Big\| \frac{1}{|{\cal T}_k|} \sum_{t_i \in {\cal T}_k} \big( {\bm x}(t_i) - \overline{\bm x}(\infty) \big) \Big\|_2^2 | {\bm x}(0) \Big\} \\
 \hfill + \mathbb{E} \Big\{ \Big\| \frac{1}{|{\cal T}_k|} \sum_{t_i \in {\cal T}_k} {\bm n}(t_i) \Big\|_2^2 \Big\}.
\end{array}
\end{equation}
It is easy to check that the latter term vanishes when $|{\cal T}_k| \rightarrow \infty$. We thus focus on the former term, which gives
\begin{equation}
\begin{array}{l}
\displaystyle \mathbb{E} \Big\{ \Big\| \frac{1}{|{\cal T}_k|} \sum_{t_i \in {\cal T}_k} \big( {\bm x}(t_i) - \overline{\bm x}(\infty) \big) \Big\|_2^2 | {\bm x}(0) \Big\} \\
\displaystyle = \frac{1}{|{\cal T}_k|^2} \mathbb{E} \Big\{ \Big\| \sum_{t_i \in {\cal T}_k} \big( \bm{\Phi}(0,t_i) - \overline{\bm W}^\infty \big) {\bm x}(0)  \Big\|_2^2 \Big\} \\
\displaystyle = \frac{1}{|{\cal T}_k|^2} \mathbb{E} \big\{ {\rm Tr} \big( \bm{\Xi} {\bm x}(0) {\bm x}(0)^T \big) \big\},
\end{array}
\end{equation}
where
\begin{equation}
\bm{\Xi} = \sum_{t_j \in {\cal T}_k} \big( \bm{\Phi}(0,t_j) - \overline{\bm W}^\infty \big)^T \sum_{t_i \in {\cal T}_k} \big( \bm{\Phi}(0,t_i) - \overline{\bm W}^\infty \big).
\end{equation}
Expanding the above product yields two groups of terms --- when $t_i = t_j$ and when $t_i \neq t_j$.
When $t_i = t_j$, using $T_o \rightarrow \infty$ and  Lemma~\ref{lemma:phi}, it is straightforward to show that:
\begin{equation} \label{eq:mat_bd}
\begin{array}{l}
\displaystyle \| \mathbb{E} \big\{ \big( \bm{\Phi}(0,t_i) - \overline{\bm W}^\infty \big)^T \big( \bm{\Phi}(0,t_i) - \overline{\bm W}^\infty \big) \big\} \| \leq C,
\end{array}
\end{equation}
for some constant $C < \infty$. As a matter of fact, we observe that the above term will not vanish at all. This is due to Observation~\ref{obs:fluctuate}, the random matrix $\bm{\Phi}(0,t_i)$ does not converge in mean square sense.

For the latter case, we assume $t_j > t_i$. We have
\begin{equation}
\begin{array}{l}
\displaystyle \big( \bm{\Phi}(0,t_j) - \overline{\bm W}^\infty \big)^T \big( \bm{\Phi}(0,t_i) - \overline{\bm W}^\infty \big) \vspace{.1cm} \\
\displaystyle = \big( \bm{\Phi}(t_i+1,t_j) \bm{\Phi}(0,t_i) - \overline{\bm W}^\infty \big)^T \big( \bm{\Phi}(0,t_i) - \overline{\bm W}^\infty \big).
\end{array}
\end{equation}
Taking expectation of the above term gives:
\begin{equation} \label{eq:inter2}
\begin{array}{l}
\displaystyle \mathbb{E} \big\{ \big( \bm{\Phi}(0,t_i) - \overline{\bm W}^\infty \big)^T \overline{\bm W}^{t_j-t_i} \big( \bm{\Phi}(0,t_i) - \overline{\bm W}^\infty \big) \big\},
\end{array}
\end{equation}
where we used the fact that $\bm{\Phi}(t_i+1,t_j)$ is independent of the other random variables in the expression and $\overline{\bm W}^{\infty} \overline{\bm W}^\ell = \overline{\bm W}^{\infty}$ for any $\ell \geq 0$.
Now, note that
\begin{equation}
\overline{\bm W}^{t_j-t_i} = \overline{\bm W}^{\infty} + {\cal O}(\lambda^{t_j-t_i}),
\end{equation}
for some $0 < \lambda \triangleq \lambda_{max}(\overline{\bm D}) < 1$. This is due to the fact that $\overline{\bm D}$ is sub-stochastic.

As $T_o \rightarrow \infty$ and by invoking Lemma~\ref{lemma:phi}, the matrix $(\bm{\Phi}(0,t_i) - \overline{\bm W}^\infty)$ has almost surely \emph{only} non-empty entries in the lower left block. Carrying out the block matrix multiplications and using the boundedless of $\bm{\Phi}(0,t_i)$ gives
\begin{equation} \label{eq:inter2}
\begin{array}{l}
\displaystyle \big\| \mathbb{E} \big\{ \big( \bm{\Phi}(0,t_j) - \overline{\bm W}^\infty \big)^T  \big( \bm{\Phi}(0,t_i) - \overline{\bm W}^\infty \big) \big\} \big\| \displaystyle \leq {\cal O}(\lambda^{t_j-t_i}).
\end{array}
\end{equation}

Combining these results, we can show
\begin{equation}
\begin{array}{l}
\displaystyle \frac{ \mathbb{E} \big\{ {\rm Tr} \big( \bm{\Xi} {\bm x}(0) {\bm x}(0)^T \big) \big\} }{|{\cal T}_k|^2} \leq \frac{{C'}}{|{\cal T}_k|} \Big( \sum_{i=0}^{|{\cal T}_k|-1} \lambda^{\min_{k} |t_{k+i} - t_k|}  \Big),
\end{array}
\end{equation}
for some $C' < \infty$. Notice that $\min_{\ell} |t_{\ell+i} - t_\ell| \geq i$ and the terms inside the bracket can be upper bounded by the summable geometric series $\sum_{i=0}^{|{\cal T}_k|-1} \lambda^i$, since $\lambda < 1$.
Consequently, the mean square error goes to zero as $|{\cal T}_k| \rightarrow \infty$. The estimator \eqref{eq:sample} is consistent.

\bibliographystyle{IEEEtran}
\bibliography{social}

\end{document}